\documentclass[fleqn,usenatbib,useAMS]{mnras}

\usepackage{newtxtext,newtxmath}
\usepackage[T1]{fontenc}
\DeclareRobustCommand{\VAN}[3]{#2}
\let\VANthebibliography\thebibliography
\def\thebibliography{\DeclareRobustCommand{\VAN}[3]{##3}\VANthebibliography}

\usepackage{graphicx}
\usepackage{amsmath}
\usepackage{deluxetable} 

\newcommand{\bea}{\texttt{BEAGLE}}
\newcommand{\Msun}{M$_{\odot}$}
\newcommand{\Msunyr}{M$_{\odot}$\,yr$^{-1}$}

\title[High-z stellar mass function]{How robustly can we constrain the low-mass end of the $z\sim6-7$ stellar mass function? -- The limits of lensing models and stellar population assumptions in the \textit{Hubble Frontier Fields}}

\author[Furtak, Atek, \& Lehnert et al.]{Lukas J. Furtak,$^{1}$\thanks{E-mail: furtak@iap.fr}
Hakim Atek,$^{1}$ 
Matthew D. Lehnert,$^{1}$
Jacopo Chevallard,$^{1}$
\newauthor{St\'{e}phane Charlot$^{1}$}\\
$^1$Sorbonne Universit\'{e}, CNRS UMR 7095, Institut d'Astrophysique de Paris, 98bis bvd Arago, 75014, Paris, France
}

\date{Accepted 2020 November 27. Received 2020 October 27; in original form 2020 June 24}

\pubyear{2020}

\begin{document}
\label{firstpage}
\pagerange{\pageref{firstpage}--\pageref{lastpage}}
\maketitle

\begin{abstract}
We present new measurements of the very low-mass end of the galaxy stellar mass function (GSMF) at $z\sim6-7$ computed from a rest-frame ultraviolet selected sample of dropout galaxies. These galaxies lie behind the six \textit{Hubble Frontier Fields} clusters and are all gravitationally magnified. Using deep \textit{Spitzer}/IRAC and \textit{Hubble Space Telescope} imaging, we derive stellar masses by fitting galaxy spectral energy distributions and explore the impact of different model assumptions and parameter degeneracies on the resulting GSMF. Our sample probes stellar masses down to $M_{\star}>10^{6}\,\text{M}_{\sun}$ and we find the $z\sim6-7$ GSMF to be best parametrized by a modified Schechter function which allows for a turnover at very low masses. Using a Monte-Carlo Markov Chain analysis of the GSMF, including accurate treatment of lensing uncertainties, we obtain a relatively steep low-mass end slope $\alpha\simeq-1.96_{-0.08}^{+0.09}$ and a turnover at $\log(M_T/\text{M}_{\sun})\simeq7.10_{-0.56}^{+0.17}$ with a curvature of $\beta\simeq1.00_{-0.73}^{+0.87}$ for our minimum assumption model with constant star-formation history (SFH) and low dust attenuation, $A_V\leq0.2$. We find that the $z\sim6-7$ GSMF, in particular its very low-mass end, is significantly affected by the assumed functional form of the star formation history and the degeneracy between stellar mass and dust attenuation. For example, the low-mass end slope ranges from $\alpha\simeq-1.82_{-0.07}^{+0.08}$ for an exponentially rising SFH to $\alpha\simeq-2.34_{-0.10}^{+0.11}$ when allowing $A_V$ of up to 3.25. Future observations at longer wavelengths and higher angular resolution with the \textit{James Webb Space Telescope} are required to break these degeneracies and to robustly constrain the stellar mass of galaxies on the extreme low-mass end of the GSMF.
\end{abstract}



\section{Introduction} \label{sec:introduction}

The study of how galaxies form and evolve during the epoch of cosmic reionization
\citep[$z\sim6-10$, ][]{fan06, planck16}, represents one of the key challenges in contemporary astrophysics. The high angular resolution and the near-infrared (NIR) sensitivity of the \textit{Hubble Space Telescope} (HST), allow us to compile statistically significant samples of high-redshift galaxies ($z\ga6$ or $<400$\,Myr after the Big Bang). Galaxies at these high redshifts have been discovered through breaks in their rest-frame ultraviolet (UV) emission due to the opacity of neutral hydrogen in the intergalactic medium (IGM). These early star-forming galaxies are of particular interest since they currently represent the best candidates for the sources responsible for reionizing the neutral IGM \citep{bunker10, oesch10a, bouwens11, bouwens15, robertson13, robertson15, mcLure13, finkelstein15, atek15b}. Using such samples of high-redshift galaxies, it is now possible to explore the build-up of stellar mass and chemical enrichment in the high-redshift ($z>6$) Universe. The galaxy stellar mass function (GSMF) can provide important constraints on theoretical models and numerical simulations of galaxy formation and evolution \citep[e.g.][]{dayal18}. Studies of the high-redshift GSMF have found much steeper low-mass slopes, $\alpha$$\sim$$-$(1.6-2.0), than those estimated at lower redshifts \citep{duncan14,grazian15,song16}. These studies, which were based on blank fields observations, are limited to galaxies with absolute UV magnitudes $M_{UV}\la-17$\,AB. Characterizing only relatively luminous galaxies means that these studies only probe stellar masses larger than $M_{\star}\sim10^{7.5}$\,\Msun\ and therefore cannot robustly constrain the very low-mass end of the GSMF.

These limits can however be pushed to fainter and lower-mass galaxies by taking advantage of the flux amplification of strong gravitational lensing (SL) offered by massive galaxy clusters \citep['cosmic telescopes';][]{maizy10,kneib11,sharon12,monna14,richard14,coe15,coe19}. The \textit{Hubble Frontier Fields} (HFF) program \citep{lotz17} obtained deep HST observations of six massive SL clusters: Abell~2744 (A2744), MACSJ0416.1-2403 (MACS0416), MACSJ0717.5+3745 (MACS0717), MACSJ1149.5+2223 (MACS1149), Abell~S1063 (S1063) and Abell~370 (A370). Additional deep ancillary observing programs with the \textit{Spitzer Space Telescope} and ground based facilities such the ESO \textit{Very Large Telescope} (VLT) make the six HFF clusters the best studied SL clusters to date and valuable fields for studying high-redshift galaxies. Indeed, using the full HFF dataset to detect gravitationally magnified $z\gtrsim6$ objects led to the discovery of the faintest high-redshift galaxies and extended their UV luminosity functions down to $M_{UV}\lesssim-13$\,AB \citep{livermore17,bouwens17a,ishigaki18,atek18}. More recently, the first measurements of $z\gtrsim6$ GSMFs using HFF HST and \textit{Spitzer} data have extended the low-mass end of the high-redshift GSMF down to $M_{\star}>10^{6}\,\text{M}_{\sun}$, confirming the steep low-mass end slopes found in blank field studies \citep{bhatawdekar19,kikuchihara20}.

Measuring stellar masses at high redshifts however remains particularly difficult for several reasons: (i) lack of robust rest-frame optical photometry for faint galaxies resulting in poorly constrained stellar masses; (ii) non-trivial systematic effects in the strength and distribution of the strong lensing caustics which significantly impact the derived intrinsic luminosity and hence the stellar mass \citep{bouwens17a,atek18}; and (iii) significant degeneracies with other physical galaxy parameters.

In this work we present a derivation of the $z\sim6-7$ GSMF, using the full data set of the HFF program and for the first time including full treatment of lensing uncertainties and the effects of missing rest-frame optical photometry. We study the impact of various SED-fitting assumptions on the resulting GSMFs in a comparative way by systematically applying different assumptions and parameter ranges in our SED-fitting analysis. Our overall goal is to assess what is believable in high-redshift GSMF studies with the data available from the current instrumentation. This paper is structured as follows: In section~\ref{sec:data}, we describe the HFF data set that we use and our methods for obtaining photometry with particular emphasis on the infrared \textit{Spitzer} data. In section~\ref{sec:SED-fitting}, we explain the SED-fitting procedure we use to derive stellar masses. We then present the resulting mass-luminosity relations in section~\ref{sec:M-Luv-relations} and our final GSMFs in section~\ref{sec:GSMF}. Finally, we discuss these results regarding photometry, lensing and SED-fitting uncertainties in section~\ref{sec:discussion} and summarize our results in section~\ref{sec:conclusion}. Throughout this paper, we adopt a standard flat $\Lambda$CDM cosmology with $H_0=70\,\frac{\text{km}}{\text{Mpc\,s}}$, $\Omega_m=0.3$ and $\Omega_{\Lambda}=0.7$. All magnitudes are quoted in the AB magnitude system \citep{oke83}.

\section{Data}\label{sec:data}

\begin{table*}
	\caption{Limiting 3$\sigma$ AB magnitudes of the 10 photometric bands used for this study. The HST limiting magnitudes were computed in \citet{atek18}, the \textit{Ks} and Spitzer limits in \citet{steinhardt20}.}
	\label{tab:field-depths}
\begin{tabular}{lcccccccccccc}
\hline
Field       &   RA          &   Dec.        &   F435W   &   F606W   &   F814W   &   F105W   &   F125W   &   F140W   &   F160W   &   \textit{Ks}      &   IRAC1   &   IRAC2\\\hline
A2744       &   00:14:21.2  &   -30:23:50.1 &   28.8    &   29.4    &   29.4    &   28.6    &   28.6    &   29.1    &   28.3    &   26.8    &   25.9    &   25.6\\
MACS0416    &   04:16:08.9  &   -24:04:28.7 &   30.1    &   29.1    &   29.2    &   29.2    &   28.8    &   28.8    &   29.1    &   26.8    &   25.9    &   26.0\\
MACS0717    &   07:17:34.0  &   +37:44:49.0 &   29.5    &   28.6    &   29.3    &   28.9    &   28.6    &   28.5    &   28.8    &   25.9\tablenotemark{a}    &   25.6    &   25.7\\
MACS1149    &   11:49:36.3  &   +22:23:58.1 &   28.6    &   28.6    &   28.6    &   28.9    &   29.3    &   29.2    &   30.1    &   26.0\tablenotemark{a}    &   25.8    &   25.6\\
S1063       &   22:48:44.4  &   -44:31:48.5 &   30.1    &   29.1    &   29.3    &   29.0    &   28.7    &   28.5    &   28.8    &   26.9    &   25.6    &   25.6\\
A370        &   02:39:52.9  &   -01:34:36.5 &   30.1    &   29.1    &   29.3    &   29.0    &   28.7    &   28.5    &   28.4    &   26.7    &   25.7    &   25.6\\\hline
\end{tabular}
\tablenotetext{a}{The \textit{Ks} band mosaics of MACS0717 and MACS1149 are shallower due to shorter exposure times for the \textit{Keck}/MOSFIRE observations \citep{brammer16}.}
\end{table*}

Observations obtained as part of the the HFF program are deep HST optical and NIR data of all six HFF clusters and their parallel fields. The clusters were observed from 2013 to 2016 in HST cycles 21 to 23 in a total of 140 orbits for each field and its parallel field. Optical data were taken with the \textit{Advanced Camera for Survey} (ACS) in three broad-band filters F435W, F606W and F814W and NIR data were taken with the \textit{Wide Field Camera Three} (WFC3) in the F105W, F125W, F140W and F160W bands. The data were reduced and drizzled into mosaics by the HFF data reduction team at the Space Telescope Science Institute\footnote{\url{http://www.stsci.edu/}} (STScI). We refer the reader to \citet{lotz17} for a detailed description of the HST data products and the data reduction pipeline. For this analysis we use the ACS mosaics generated using the 'self-calibrating' method and the WFC3/IR mosaics that were corrected for time-variable sky background, available in the \texttt{Mikulski Archive for Space Telescopes}\footnote{\url{https://archive.stsci.edu/pub/hlsp/frontier}} (\texttt{MAST}).

In addition to HST observations, $\sim$1000 hours of Director's Discretionary time on \textit{Spitzer} were dedicated to observing the HFF clusters with the \textit{Infrared Array Camera} \citep[IRAC;][]{fazio04}. These observations had total integration times of $\sim$50 hours each in \textit{Spitzer}/IRAC channel 1 ($3.6$\,\micron) and channel 2 ($4.5$\,\micron) and resulted in depths of $\sim$26\,mag -- deep enough to detect (some) high-redshift galaxies. \textit{Spitzer}/IRAC observations at $>$3\,\micron~are crucial to the study of galaxies at $z\gtrsim6$ because IRAC channels 1 and 2 provide photometry at rest-frame optical wavelengths which are red-ward of the Balmer-/4000\,\AA~break. We therefore also use \textit{Spitzer}/IRAC channel 1 ($3.6$\,\micron) and channel 2 ($4.5$\,\micron) mosaics from the \textit{Spitzer Frontier Fields} program \citep{lotz17} in our analysis (hereafter referring to the two IRAC channels as IRAC1 and IRAC2).

The HFF clusters were also observed in the \textit{Ks}-band centered around $2.2$\,\micron, to sufficient depth to provide an additional photometric band that fills the gap between the reddest HST band and the two IRAC channels used in our study. \textit{Ks}-band data were obtained for the southern clusters with the \textit{High Acuity Wide field K-band Imager} (HAWK-I) on the ESO/VLT and in somewhat shorter integration times with MOSFIRE on the \textit{Keck} telescope of the northern clusters \citep{brammer16}. We use \textit{Ks}-mosaics drizzled to the HST 0.06\arcsec pixel-scale in this study \citep{shipley18}.

All 10 broad-band filters used in our analysis and their 3$\sigma$ limiting magnitudes can be found in Table~\ref{tab:field-depths}. We use limiting magnitudes computed in \citet{atek18} for the HST bands and in \citet{steinhardt20} for the \textit{Ks} and IRAC bands.

\subsection{HST photometry and Dropout selection} \label{sec:hst-data}

We use the $z\sim6-7$ sample detected in the six HFF clusters and the HST photometry presented in \citet{atek18}. This section summarizes the detection and measurement methods and we refer the reader to \citet{atek15a,atek18} for a detailed description of the procedure.

All HST frames were convolved to the same Point Spread Function (PSF) of the F160W frame using a PSF model computed with the \texttt{Tinytim} tool \citep{krist11}. Sources were extracted using the \texttt{SExtractor} software \citep{bertin96} in dual mode on a weighted and intra-cluster-light (ICL) corrected stack of the four WFC3/IR bands as detection image and the original frames in each band as measurement images for the photometry. ICL-correction was performed on the detection stack with a $2\arcsec\times2\arcsec$ median filter and the \texttt{SExtractor} parameters were optimized for detecting the faintest sources.
The $z\sim6-7$ sample is color-color selected using the isophotal magnitudes (\texttt{MAG\_ISO}) estimated by \texttt{SExtractor}. The selected $z\sim6-7$ candidates satisfy the Lyman break criteria \citep[see e.g.][]{steidel96} which are described in detail in \citet{atek15a}:

\begin{equation} \label{eq:dropout-criteria}
\begin{split}
    (I_{814}-Y_{105})&>1.0\\
    (I_{814}-Y_{105})&>0.6+2.0(Y_{105}-J_{125})\\
    (Y_{105}-J_{125})&<0.8
\end{split}
\end{equation}

All sources must also be detected above the $5\sigma$ level in at least two WFC3/IR bands and at $6.5\sigma$ in the NIR detection stack and satisfy a \textit{non}-detection criterion in the blue F435W and F606W bands and a stack of the two. A visual inspection of each source eliminates any remaining spurious detections and sources that have sizes that are below that expected for a point source (i.e., smaller than that of the point spread function of the image stacks given the signal-to-noise of the source). The final $z\sim6-7$ galaxy sample we use contains 303 sources. Photometric redshifts were computed with \texttt{Hyperz} \citep{bolzonella11} by \citet{atek18}.

\subsection{\textit{Ks} and \textit{Spitzer}/IRAC photometry} \label{sec:ir-Data}

\begin{table}
	\caption{Aperture diameters and aperture correction factors used for the \textit{Ks} and the two IRAC bands.}
	\label{tab:apertures}
\begin{tabular}{lccccc}
\hline
Band                             &  pixel-scale     &   $D_{ap}$    &   $D_{in}$    &   $D_{out}$   &   $c$\\\hline
\textit{Ks} (HAWK-I)\tablenotemark{a}     &  0.06\arcsec/pix &   0.5\arcsec  &   1.0\arcsec  &   1.5\arcsec  &   1.98\\
\textit{Ks} (MOSFIRE)\tablenotemark{b}    &  0.06\arcsec/pix &   0.5\arcsec  &   1.0\arcsec  &   1.5\arcsec  &   2.77\\
IRAC1           &   0.6\arcsec/pix    &   1.7\arcsec  &   2.0\arcsec  &   3.7\arcsec  &   3.35\\
IRAC2           &   0.6\arcsec/pix    &   1.3\arcsec  &   2.0\arcsec  &   3.3\arcsec  &   4.81\\\hline
\end{tabular}
\tablenotetext{a}{For A2744, MACS0416, S1063 and A370}
\tablenotetext{b}{For MACS0717 and MACS1149}
\end{table}

\begin{figure*}
    \centering
    \includegraphics[width=\textwidth, keepaspectratio=True]{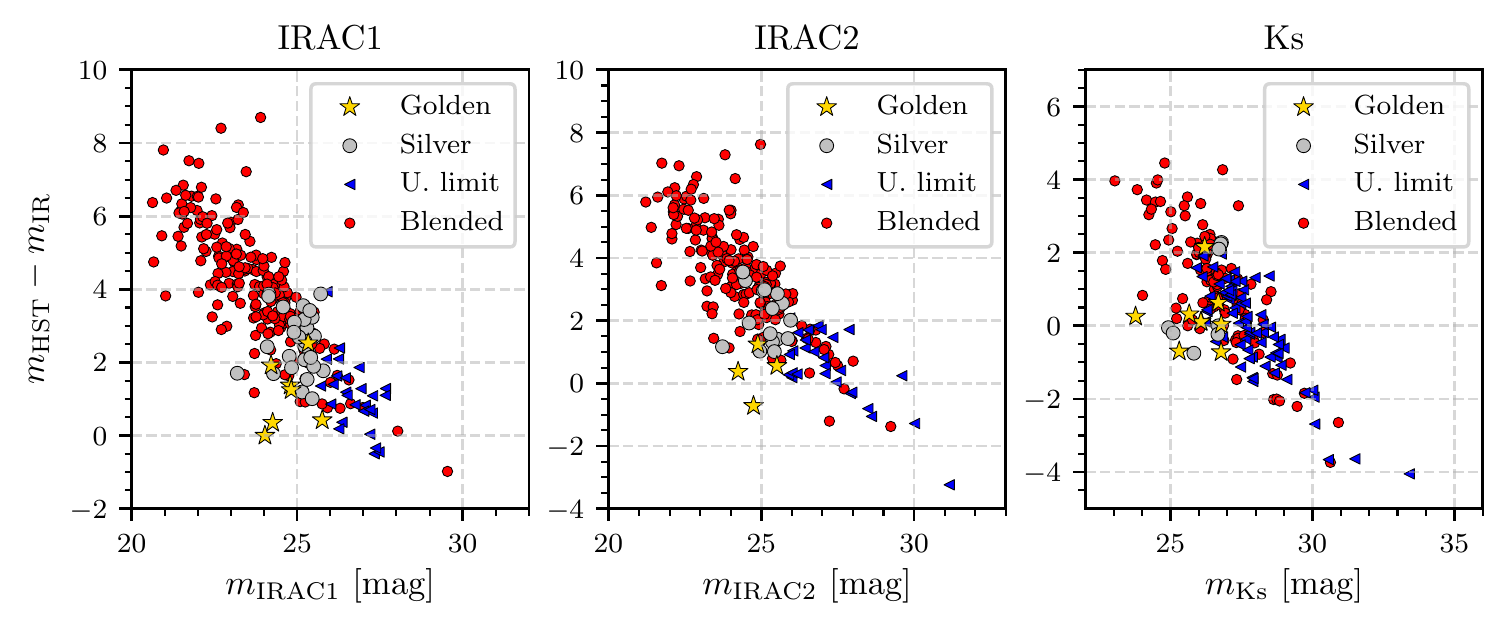}
    \caption{Observed magnitude differences with the reddest HST filter (F160W) as a function of IR magnitude in each IR band. Sources $\gtrsim4$ magnitudes brighter than F160W in IRAC1 and IRAC2 and $\gtrsim2$ magnitudes brighter in \textit{Ks} are all contaminated by bright foreground objects (red dots). Sources that are not contaminated but lie beyond the $3\sigma$ detection limits (listed in Table~\ref{tab:field-depths}) are shown as blue triangles. Note that in the \textit{Ks} band (far right panel) the main issue is not blending with foreground sources but rather the fact that most sources are undetected above $3\sigma$.}
    \label{fig:IRAC-diagnostics}
\end{figure*}

The \textit{Ks} and \textit{Spitzer}/IRAC data have a much lower spatial resolution than the HST images in which our high-redshift objects are detected. Obtaining accurate infrared photometry of high-redshift objects therefore requires great care, especially in crowded fields like the HFF clusters' centers where our very faint sources are likely to be blended with brighter foreground galaxies.

In order to estimate the stellar masses, we need to consider how our aperture photometry is related to the total flux. We compute aperture correction factors using measured PSFs from the \texttt{HFF-DeepSpace} catalogs \citep{shipley18} for the \textit{Ks} band and from the \textit{Spitzer}/IRAC handbook\footnote{\url{https://irsa.ipac.caltech.edu/data/SPITZER/docs/irac/calibrationfiles/psfprf/}} (warm mission) for the IRAC bands. Our aperture correction factors can be found in Table~\ref{tab:apertures}. We use the \texttt{photutils v0.7.2} package \citep{bradley19} to measure photometry in circular apertures whose diameters are optimized to neither overestimate the flux of small sources nor underestimate the flux of large sources. The local background for every source is measured as the mean flux in a circular annulus and then subtracted from the source flux. The diameters of the apertures and their annulii are also shown in Table~\ref{tab:apertures} for each band.

Since our sample of $z\sim6-7$ galaxies is very faint and their stellar masses are primarily constrained by their (rest-frame optical) IRAC photometry, we must carefully assess the their quality. We therefore visually inspected each source in the HST F125W band, the \textit{Ks} band, and the two IRAC $3.6$\,\micron~and $4.5$\,\micron~bands (referred to as IR bands hereafter) specifically to determine how blended it is with the light profiles of any foreground galaxies. We then assign flags to the IR photometry of each source as follows: 

\begin{itemize}
    \item \textit{Golden} source: Completely isolated source in a blank area of the field without any significant contamination from the light profile of foreground galaxies.
    \item \textit{Silver} source: Trusted photometry, i.e., there is no other source within this source's aperture or background annulus. There are however other galaxies or ICL in close proximity which might affect its photometry.
    \item \textit{Blended} source: Contaminated photometry, i.e., there is obvious light from another galaxy or star in the source's aperture or background annulus.
\end{itemize}

We make the distinction between golden and silver sources in order to have a sample of galaxies that can be trusted to have absolutely clean IRAC photometry since determining the impact of having IRAC photometry for some sources and not for others is a major focus of this study. Note however that the majority of these isolated sources are either too faint to be detected in the IRAC bands or subject to very high gravitational magnification factors and therefore large lensing uncertainties, as we will address subsequently in sections~\ref{sec:SL} and \ref{sec:IRAC-photometry_correction}. We do not use the IR photometry of blended sources in our analysis from here on.

We show the difference between observed IR magnitudes and the reddest HST band (F160W) as a function of IR magnitude for our $z\sim6-7$ sample in Fig.~\ref{fig:IRAC-diagnostics}. As can be seen, we have a significant number of sources with unrealistically high infrared fluxes, i.e. $m_{F160W}-m_{IRAC}\geq4$ and $m_{F160W}-m_{Ks}\geq2$ which are all blended with other objects (Fig.~\ref{fig:IRAC-diagnostics}). Many sources in our sample are fainter than the $3\sigma$-limiting magnitudes in the IR bands (blue triangles in Fig.~\ref{fig:IRAC-diagnostics}). We expected this result since many of our sources are at the very limit of HST detectability. These sources only have an upper limit on their IR fluxes.
Out of our sample of 303 high-redshift sources in the six HFF fields only 35 (12\,\%) have useful, i.e. golden or silver, IRAC1 photometry, 20 (7\,\%) have useful IRAC2 photometry and only 17 (6\,\%) have useful \textit{Ks} photometry. In total 39 (13\,\%) of our sources have useful photometry in either IRAC1 or IRAC2 and we discard the IR photometry of blended sources entirely. In the following sections, we discuss how SED-fitting using only HST data affects the stellar mass results.

\section{SED-fitting} \label{sec:SED-fitting}

To obtain stellar masses and other galaxy parameters, we fitted spectral energy distribution (SED) templates to the photometry of our $z\sim6-7$ sample using the \texttt{BayEsian Analysis of GaLaxy sEds (BEAGLE)} tool \citep{chevallard16}. \texttt{BEAGLE} is optimized to accurately estimate both redshift and physical galaxy parameters using a Bayesian Monte-Carlo Markov-Chain (MCMC) analysis. By performing a Bayesian MCMC analysis on each SED fit, \texttt{BEAGLE} rigorously probes a large parameter space and efficiently quantifies parameter uncertainties. Its modular design allows for maximum flexibility in changing the assumptions and models that underlie the fits such as the stellar population models, nebular emission templates, dust attenuation laws, physical parameters of the models, etc. We take advantage of this feature of \texttt{BEAGLE} to probe the impact of various assumptions commonly made in this type of photometric study on the resulting stellar masses (section~\ref{sec:assumptions}).

SED-fitting of high-redshift galaxies has been found to strongly depend on nebular emission as well as stellar population models. Several studies have shown that SEDs including nebular emission fit such objects significantly better and yield lower stellar masses and ages \citep[e.g][]{schaerer09,schaerer10,ono10,atek11,mclure11,duncan14}. We therefore adopt stellar and nebular SED templates computed by \citet{gutkin16} which combine the latest version of the \citet{bc03} stellar population models with the photoionization code \texttt{CLOUDY} \citep{ferland13}. \texttt{BEAGLE} then includes the latest analytical IGM absorption models by \citet{inoue14} and applies a dust attenuation law to the galaxy templates in order to account for dust attenuation within the fitted galaxy.

We use all ten bands of broad-band photometry in the SED-fitting analysis if available. The total magnitudes were derived using \texttt{SExtractor} and are estimated using \texttt{MAG\_AUTO} for the HST images. The \textit{Ks} and IRAC total magnitudes were measured as described in section~\ref{sec:ir-Data} and are only used for golden and silver sources. We estimate upper limits for all sources that are not detected at the $3\sigma$ level in any of the images. We furthermore fix the galaxies' redshift to the values obtained by \citet{atek18} with \texttt{Hyperz} throughout our analysis (see section~\ref{sec:hst-data}).

\subsection{SED-fitting assumptions} \label{sec:assumptions}

\begin{table}
	\caption{Fit parameters and priors of our SED-fitting analysis with \texttt{BEAGLE}. The \textit{Reference model} represents the most basic configuration used here. Parameters that depart significantly from the reference model configuration are also listed.}
	\label{tab:SED-fit_parameters}
\begin{tabular}{lccc}
\hline
Parameter                   &   Unit                    &   Prior   &   Value/range\\\hline
\multicolumn{4}{c}{\textit{Reference model}}\\
\hline
$\log M_{\star}$            &   $\log \text{M}_{\sun}$  &   flat    &   [5.0, 11.0]\\
$\log t_{\text{age}}$       &   $\log \text{yr}$        &   flat    &   [7.37, 8.37]\\
$\hat{\tau}_V$              &   -                       &   flat    &   [0.0, 0.2]\\
$\log\hat{U}$               &   -                       &   flat    &   [-3.0, -1.0]\\
$Z$                         &   $\text{Z}_{\sun}$       &   fixed   &   0.1\\
SFH                         &   -                       &   fixed   &   $\psi(t)=\psi$\\
Dust attenuation law            &   -                       &   fixed   &   Calzetti\\
\hline
\multicolumn{4}{c}{\textit{Metallicity tests}}\\
\hline
$Z$                         &   $\text{Z}_{\sun}$       &   fixed   &   0.01, 0.1, 0.5\\  
\hline
\multicolumn{4}{c}{\textit{SFH tests}}\\
\hline
                            &                           &           &   $\psi(t)\propto t\exp{(-t/\tau})$\\
SFH                         &   -                       &   fixed   &   $\psi(t)\propto\exp({-t/\tau})$\\
                            &                           &           &   $\psi(t)\propto\exp({t/\tau})$\\
$\log \tau$                 &   $\log \text{yr}$        &   flat    &   [7.0, 8.5]\\
\hline
\multicolumn{4}{c}{\textit{Dust tests}}\\
\hline
$\hat{\tau}_V$              &   -                       &   flat    &   [0.0, 3.0]\\
Dust attenuation law        &   -                       &   fixed   &   Calzetti, SMC\\
\hline
\end{tabular}
\par\bigskip
Note. -- $M_{\star}$: Stellar mass -- $t_{\text{age}}$: Maximum stellar age -- $\hat{\tau}_V$: \textit{V}-band attenuation optical depth -- $\hat{U}$: Effective galaxy-wide ionization parameter -- $Z$: Stellar metallicity -- $\tau$: Star formation $e$-folding time
\end{table}

The best-fit SED depends on numerous parameters, namely star-formation history (SFH), metallicity, dust attenuation, stellar age and star-formation timescales. Moreover, several parameters, such as e.g., stellar mass and stellar age, are degenerate. In order to obtain accurate stellar masses we need to make a number of assumptions regarding these parameters. However, the physical properties and star formation histories of high-redshift galaxies remain largely unknown. We therefore use \texttt{BEAGLE}'s modular build to test the impact of these assumptions on the main parameter of interest: stellar mass and the resulting stellar mass function at $z\sim6-7$.

This is done by first defining a \textit{reference model}, a configuration of \texttt{BEAGLE} fit-parameters from which other SED-fit runs depart to explore the impact of different parameter assumptions on stellar mass. The reference model is defined as the most basic "minimal" configuration possible to fit stellar mass with four free parameters: Stellar mass $M_{\star}$, maximum stellar age $t_{\text{age}}$, effective \textit{V}-band dust optical depth $\hat{\tau}_V$, and the effective galaxy-wide ionization parameter $\hat{U}$. All other parameters are fixed: We assume a constant SFH, a metallicity of $0.1\,\text{Z}_{\sun}$, and a Calzetti dust attenuation law \citep{calzetti94,calzetti00}. All parameters and their priors are summarized in Table~\ref{tab:SED-fit_parameters}.

The allowed ranges of $t_{\text{age}}$ and $\hat{\tau}_V$ are carefully chosen to limit degeneracy with stellar mass. The lower and upper bounds on the range of stellar age allowed in the SED modeling represent, respectively, a rough estimate of a galaxies' dynamical timescale given its typical size and stellar mass within the sample of galaxies, $t_{\text{dyn}}$ and $10\times t_{\text{dyn}}$. The dynamical time $t_{\text{dyn}}$ is estimated as,

\begin{equation} \label{eq:t_dyn}
    t_{\text{dyn}}\sim\frac{r}{v}\sim\sqrt{\frac{2r^3}{GM}}
\end{equation}

\noindent
where $r$ is a characteristic size of the galaxy and $v$ the virial velocity \citep{verma07}. We use $r=0.5$\,kpc, a typical half-light radius of a galaxy at $z\sim6-7$ \citep{kawamata15}, and $M\sim10^{8}\,\text{M}_{\sun}$ as a typical galaxy mass, resulting in a dynamical time $t_{\text{dyn}}\sim20$\,Myr. We set the lower boundary of $t_{\text{age}}$ to 20\,Myr in order to avoid solutions around $t_{\text{age}}\sim10$\,Myr with extremely high nebular equivalent widths (EWs) $\sim7000$\,\AA~for [\ion{O}{iii}]\,5007\,\AA, which remain unlikely given evidence from observations of high-redshift galaxies \citep{atek11,atek14b,smit14,reddy18b} and stellar population models \citep[but see][]{endsley21}. We refer the reader to appendix~\ref{app:age-impact} for further discussion of this issue. Dust attenuation is allowed to vary between $\hat{\tau}_V=0$, i.e., no dust attenuation, and $\hat{\tau}_V=0.2$ which corresponds to $A_V\approx0.2$ and $A_{UV}\approx0.8$. We chose this upper boundary because high-redshift galaxies are unlikely to have high dust attenuation \citep[e.g.,][]{bouwens16}.

To gauge the influence of specific assumptions in our reference set of priors and values (when they are fixed), we ran different SED fitting configurations with \bea\ where we change only one or two parameters or priors at a time to study their impact on the resulting stellar mass function. The various configurations we tried are listed in Table~\ref{tab:SED-fit_parameters}. We ran tests to determine the impact of increasing the latitude in the range of the parameters allowed and through changing the functional dependence of parameters of the metallicity, SFH and dust attenuation. We summarize the changes we made in the assumptions in SED fitting as:

\begin{itemize}
    \item \textit{Metallicity}: To determine the impact of different assumptions for the metallicity, we ran models with the same configuration as the reference configuration except fixing the metallicity, $Z$, to three different values: 0.01\,Z$_{\sun}$, 0.1\,Z$_{\sun}$ and 0.5\,Z$_{\sun}$, the last of which corresponds to the metallicity measured from absorption spectra of a small sample of bright (i.e. massive) galaxies at $z\sim6$ \citep{harikane20}.
    
    \item \textit{SFH}: We determined the impact of the functional changes in the SFH by running models with three additional functional forms of the SFH: a delayed SFH $\psi(t)\propto t\exp{(-t/\tau)}$ and exponentially rising $\psi(t)\propto\exp{(t/\tau)}$ or declining $\psi(t)\propto\exp{(-t/\tau)}$. This approach is similar to that of \citet{grazian15} who tested these three SFHs for galaxies at $3.5\leq z\leq4.5$. We note though that they did not consider the impact of nebular emission which we do. These SFHs require an additional free parameter, the star-formation $e$-folding time, $\tau$. The delayed SFH has a peak of star-formation at $t=\tau$.
    
    \item \textit{Dust}: We tested the impact of dust attenuation on the results by allowing a wider range of dust attenuation optical depths in the \textit{V}-band, $\hat{\tau}_V$, compared to that allowed in the reference model (up to $\hat{\tau}_V=3$ or $A_V=3.25$).  This is similar to the range allowed in \citet{song16,bhatawdekar19,kikuchihara20}. Recent results have also indicated that a steeper dust attenuation law than the Calzetti law potentially fits better the SEDs of high-redshift galaxies \citep{capak15,reddy15,reddy18a}. We therefore also ran a model where we assume an SMC-like extinction law \citep{pei92} with a wide range of allowed attenuation optical depths, $\hat{\tau}_V\in[0, 3]$, in fitting the SED.
\end{itemize}

\subsection{SL magnification} \label{sec:SL}

\begin{figure}
    \centering
    \includegraphics[width=0.5\textwidth, keepaspectratio=True]{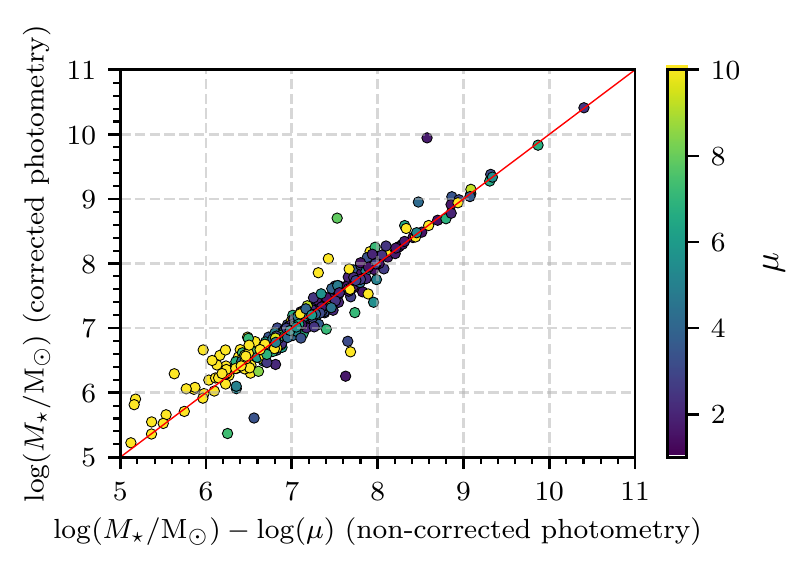}
    \caption{A comparison between the stellar masses determined by fitting the SEDs using the observed fluxes with those determined by fitting magnification corrected photometry. The color-coding indicates the median magnification factor, $\mu$, from the six SL models considered in this analysis and its values are shown in the color bar on the right.}
    \label{fig:mu-correction_legtimiation}
\end{figure}

In computing rest-frame UV luminosities and stellar masses, we need to determine gravitational magnification of each galaxy. Several independent teams produced numerous SL models for all six HFF clusters, which are all publicly available in the \texttt{MAST} archive.\footnote{see e.g., \url{https://archive.stsci.edu/pub/hlsp/frontier/abell2744/models/}} In this study we restrict ourselves to models that are both available for all six clusters and are based on the full and most recent data sets available from the HFF program: CATS \citep{jauzac14,jauzac15,limousin16,lagattuta17}, Diego \citep{diego15}, GLAFIC \citep{kawamata16,kawamata18}, Keeton \citep{ammons14,mccully14}, Sharon \& Johnson \citep{johnson14} and Williams \citep{grillo15}.

We compute gravitational magnification factors at each galaxy's position and photometric redshift from the convergence $\kappa$ and shear $\gamma$ maps provided by the SL modelling teams as,

\begin{equation} \label{eq:mu}
	\mu=\frac{1}{(1-\kappa)^2-\gamma^2}.
\end{equation}

\noindent
The resulting magnification factors can differ significantly between the models due to different SL modelling techniques and assumptions \citep[e.g.,][]{meneghetti17,acebron17,acebron18}. These differences can significantly affect any deductions made about the luminosity and mass functions of high-redshift galaxies \citep{bouwens17a,atek18}. In order to take these differences into account and to assess the systematic lensing uncertainties, we compute the median magnification factor, $\mu$, and the standard deviation between all the models, $\Delta\mu$, and assume the standard deviation is the uncertainty in the lensing strength for each galaxy. We will further discuss the impact of using this estimate as the uncertainties on the lensing strength in section~\ref{sec:gravlens-discussion}.

The best-fit stellar masses are directly corrected for median magnification as determined from the different lensing models. The validity of this approach is shown in Fig.~\ref{fig:mu-correction_legtimiation}: Stellar masses corrected \textit{a posteriori} for the median magnification widely agree with masses fit to magnification-corrected photometry. It also confirms that the galaxies with the lowest masses correspond to those which are most strongly magnified.

\subsection{Correcting for missing IRAC photometry} \label{sec:IRAC-photometry_correction}

\begin{figure}
    \centering
    \includegraphics{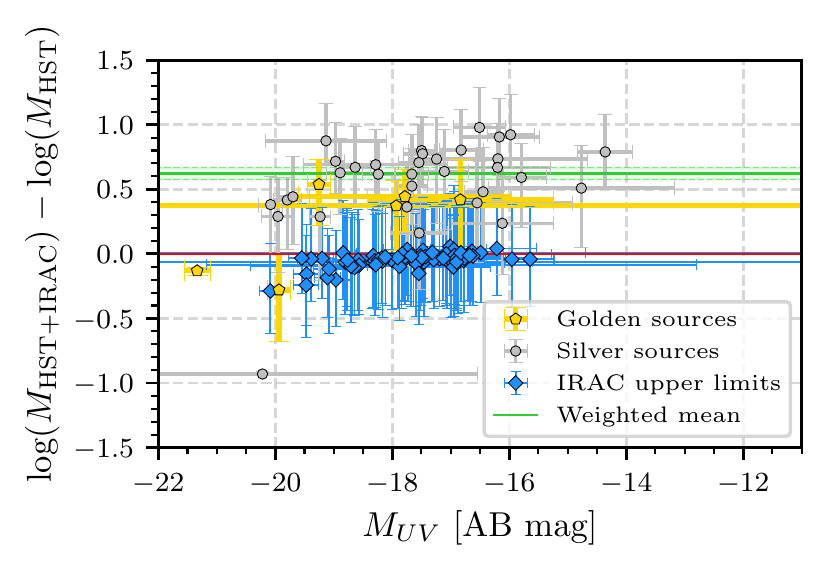}
    \caption{Difference in stellar mass derived from fitting HST+IRAC photometry and HST photometry only as a function of UV luminosity, $M_{UV}$ (the absolute magnitude at 1500\,\AA), using the reference model \texttt{BEAGLE}-parameters. Most of the golden and silver sources (cf. text for details) lie above the zero level, indicating that fitting only the HST data significantly underestimates the stellar mass. Sources with upper limits in the flux in both of the IRAC bands (blue squares), lie close to zero, which means that upper limits on IRAC flux are not sensitive enough to provide any additional constrains on the stellar mass. The average correction offset, $\langle\delta\rangle=0.62\pm0.05~\text{dex}$, is shown as the green line and the shaded region indicates its 1$\sigma$ uncertainties.}
    \label{fig:IRAC-photometry_correction}
\end{figure}

As discussed in Section~\ref{sec:ir-Data}, the vast majority of our sources, 87\%, do not have reliable \textit{Spitzer}/IRAC photometry in at least one band either because they are severely blended (i.e., there is no information), have unreliable photometry due to possible contamination, or are undetected by the $3\sigma$ limits. However, rest-frame optical photometry is crucial for making more robust estimates of stellar masses at $z\sim6-7$. We only fit the 7 bands of HST for most of our sources and therefore we only have rest-frame UV photometry, which underestimates the stellar mass of these galaxies. We now quantify how much we are underestimating the stellar masses of galaxies when only using rest-frame UV photometry.

In order to estimate the magnitude and difference in using only the HST photometry versus using HST plus IRAC photometry in the final stellar mass, we fitted the SEDs of sources with and without including the IRAC photometry. For this comparison, we used the same set of parameter values when fitting the SEDs and included all golden and silver sources, and robust upper limits in the IRAC photometry. We find that fitting only the HST data underestimates the stellar mass by $\gtrsim0.5$~dex (Fig.~\ref{fig:IRAC-photometry_correction}). Note that sources that are not detected at the $3\sigma$-level in IRAC1 \textit{and} IRAC2, i.e. the blue squares in Fig.~\ref{fig:IRAC-photometry_correction}, all lie close to zero in the ordinate. This implies that there is no significant difference in the stellar mass estimate when fitting upper limits in the IRAC bands compared to only fitting the HST photometry. We therefore conclude that the upper limits in the IRAC photometry are not sensitive enough to provide robust constraints on the stellar mass (cf. also Fig.~\ref{fig:IRAC-photometry_SEDs} in appendix~\ref{app:irac-correction_bias}).

There are three outliers below zero among the golden and silver sources in Fig.~\ref{fig:IRAC-photometry_correction} with bright rest-frame UV luminosities $M_{UV}\lesssim-20$. An inspection of their best-fit SEDs reveals a poor fit due to the relatively low upper boundary on dust attenuation used in the reference model (cf.~Table~\ref{tab:SED-fit_parameters}). These galaxies require a larger range on $\hat{\tau}_V$ to accurately fit their SED. Since these three galaxies lie at the most luminous end of the $M_{UV}$ range that we probe, we do not consider them representative of our entire sample however.

We compute the mean $\delta=\log(M_{\mathrm{HST+IRAC}}/\text{M}_{\sun})-\log(M_{\mathrm{HST}}/\text{M}_{\sun})$ from all sources with trusted IRAC photometry (i.e., golden and silver sources), excluding the outliers. We find a correction factor of $\langle\delta\rangle\simeq0.62\pm0.05~\text{dex}$ (cf.~Fig.~\ref{fig:IRAC-photometry_correction}). Since this offset appears to be constant over the range of rest-frame UV luminosities that we probe in this work (cf. Fig.~\ref{fig:IRAC-photometry_correction_fit} in appendix~\ref{app:irac-correction_bias}), we apply this correction to the stellar mass of every galaxy which is either contaminated or only has upper limits in its IRAC photometry. Note that as the galaxies for which we compute the correction $\langle\delta\rangle$ are well detected in at least one of the IRAC bands, they represent a rest-frame optically bright population with relatively high stellar masses given their UV luminosities. We would therefore expect our correction factor to slightly overestimate the stellar mass of the lower-mass galaxies in our sample which would not be IRAC detected (cf.~appendix~\ref{app:irac-correction_bias})

\section{Mass-Luminosity scaling relation} \label{sec:M-Luv-relations}

\begin{figure}
    \centering
    \includegraphics{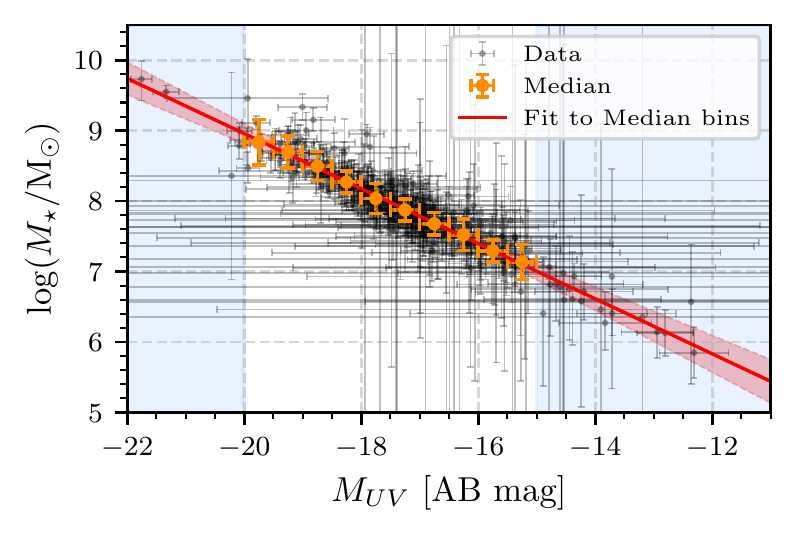}
    \caption{Mass-luminosity relation for the reference \texttt{BEAGLE} configuration (Table~\ref{tab:SED-fit_parameters}). Black dots represent individual galaxies and their uncertainties in $M_{\star}$ and $M_{UV}$. The uncertainties are dominated by the gravitational magnification at $M_{UV}\gtrsim-17$. Orange dots represent the median-mass bins of 0.5\,mag width used for the fit. The best-fit mass-luminosity relation is shown as a red line and its 1$\sigma$-range as the red shaded area. The blue shaded areas represent the regions where the luminosity bins contain too few objects to robustly constrain the shape of the $M_{\star}-M_{UV}$-relation. We exclude all bins in these two regions from the fit.}
    \label{fig:m-muv-relation_data}
\end{figure}

\begin{figure*}
    \centering
    \includegraphics[width=\textwidth, keepaspectratio=True]{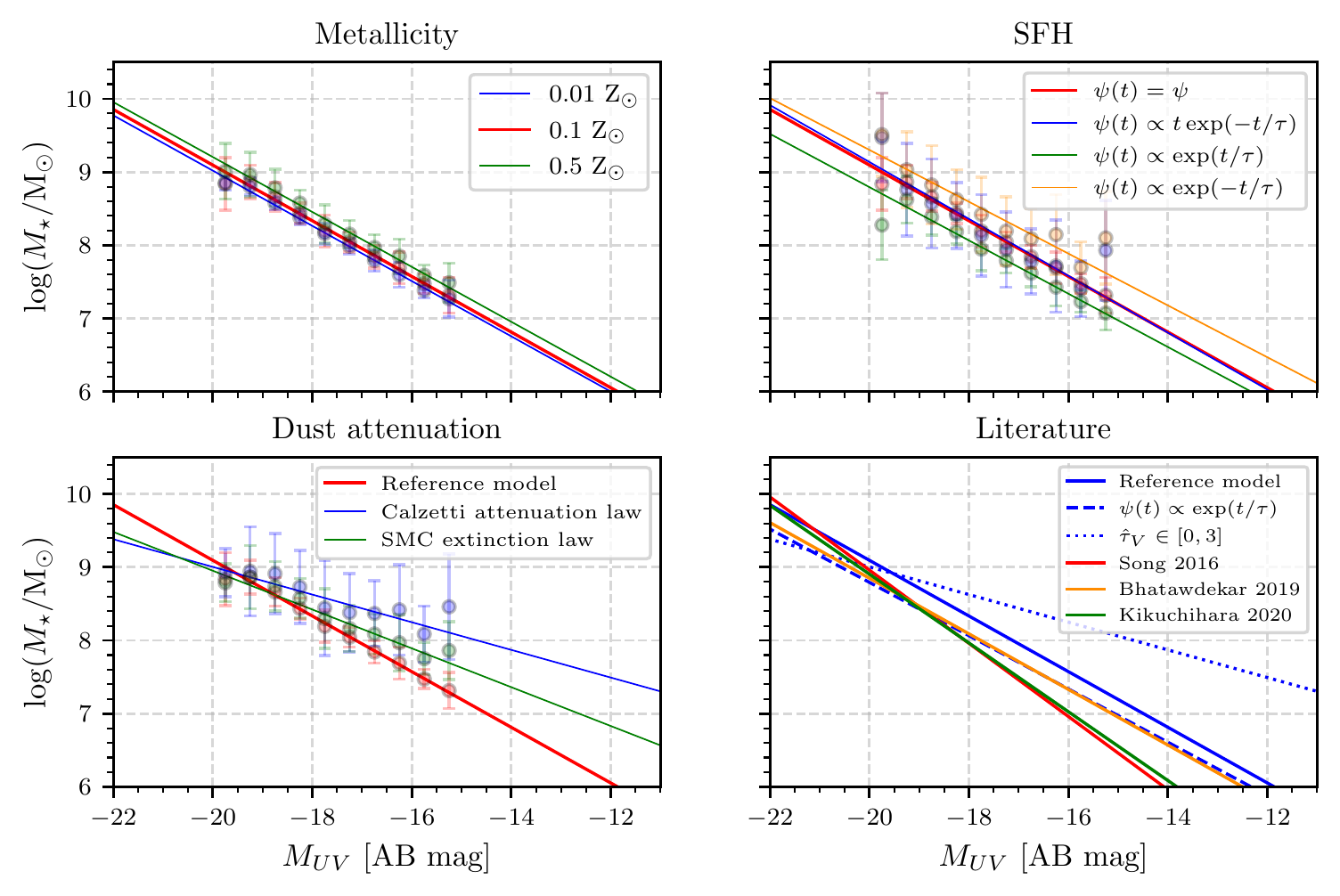}
    \caption{Best-fit $M_{\star}-M_{UV}$-relations for the different \texttt{BEAGLE} configurations as discussed in section~\ref{sec:assumptions}. {\it (Upper left panel)}: Best-fit $M_{\star}-M_{UV}$-relations for three different metallicities; {\it (Upper right panel)}: for the four different SFHs; and {\it (Lower left panel)}: for three different attenuation laws (all the various assumptions and differences are indicated in the legend in each panel). The data points indicate the median of the mass bins used for the fit and the solid lines represent the median best-fit relations determined from our MCMC analysis. {\it (Lower right panel)}: We compare our results for our reference model, an exponentially rising SFH, and a wide range of attenuation values (blue solid, dashed and dotted lines respectively).  In addition we show the relations determined in other recent studies of high-redshift galaxies: \citet{song16} (red line), \citet{bhatawdekar19} (orange line) and \citet{kikuchihara20} (green line). Note that while \citet{song16} and \citet{kikuchihara20} find steeper relations than ours and that of \citet{bhatawdekar19}, all relations yield similar stellar masses at the bright end, indicating that the stellar masses for these galaxies are constrained within a factor of $\sim$0.5\,dex.}
    \label{fig:m-muv-relations}
\end{figure*}

\begin{table}
	\caption{Parameters of fits to the mass-luminosity relation given in Eq.~\ref{eq:m-muv-relation}. The best-fit parameter values are medians and their $1\sigma$-uncertainties are drawn from our MCMC analysis.}
	\label{tab:M-Muv-Relations}
\begin{tabular}{lccc}
\hline
\texttt{BEAGLE}                     &   Slope           &   Intercept       &   $\log M_{\star}$\\
Configuration                       &                   &                   &   \scriptsize{($M_{UV}=-$19.5)}\\
                                    &                   &   (M$_{\sun}$)    &   (M$_{\sun}$)\\
\hline
Reference Model                     &   $-0.38\pm0.05$  &   $-0.05\pm0.06$  &   $8.90\pm0.06$\\
\hline
$Z=0.01\,\text{Z}_{\sun}$           &   $-0.38\pm0.03$  &   $-0.11\pm0.05$  &   $8.83\pm0.02$\\
$Z=0.5\,\text{Z}_{\sun}$            &   $-0.37\pm0.05$  &   $-0.08\pm0.07$  &   $9.01\pm0.07$\\
\hline
$\psi(t)\propto t\exp{(-t/\tau)}$   &   $-0.39\pm0.12$  &   $-0.03\pm0.17$  &   $8.94\pm0.13$\\
$\psi(t)\propto\exp{(t/\tau)}$      &   $-0.36\pm0.05$  &   $-0.30\pm0.07$  &   $8.61\pm0.07$\\
$\psi(t)\propto\exp{(-t/\tau)}$     &   $-0.35\pm0.11$  &   $0.24\pm0.15$   &   $9.12\pm0.12$\\
\hline
Calzetti attenuation law            &   $-0.19\pm0.10$  &   $0.44\pm0.17$   &   $8.91\pm0.08$\\
SMC-like extinction law             &   $-0.27\pm0.07$  &   $0.16\pm0.10$   &   $8.82\pm0.06$\\
\hline
\end{tabular}
\end{table}

In order to determine the scaling relation between the absolute rest-frame UV magnitude $M_{UV}$ and stellar mass $M_{\star}$, we estimate the mass-to-light ratio of our galaxy sample. The rest-frame UV magnitude is defined as the rest-frame luminosity at 1500\,\AA. At the redshifts studied here, 1500\,\AA~is redshifted into the HST \textit{Y$_{105}$} and \textit{J$_{125}$} bands. In determining this mass-to-light ratio for each source, we included the median magnification from the SL models and the dispersion between models in the final total uncertainties. In Fig.~\ref{fig:m-muv-relation_data}, we present the stellar masses obtained for the reference \texttt{BEAGLE} configuration (see section~\ref{sec:assumptions} and Table~\ref{tab:SED-fit_parameters}) as a function of $M_{UV}$. As expected, we observe a strong correlation between stellar mass and UV luminosity. At the faint end ($M_{UV}\gtrsim-17$), the large uncertainties of the gravitational magnification estimates (see section~\ref{sec:SL}) begin to dominate the uncertainties in both $M_{\star}$ and $M_{UV}$. The data are binned into $M_{UV}$ bins of 0.5\,mag width and the median $M_{\star}$ is taken in each luminosity bin (orange dots in Fig.~\ref{fig:m-muv-relation_data}). The error bars account for the scatter in stellar mass in each bin. We then fit the relation,

\begin{equation} \label{eq:m-muv-relation}
    \log\left(\frac{M_{\star}}{\mathrm{M}_{\sun}}\right)-8=a(M_{UV}+17)+b
\end{equation}

\noindent
to the median bins. We exclude luminosity bins fainter than $M_{UV}>-15$ and  brighter than $M_{UV}<-20$ (blue shaded area in Fig.~\ref{fig:m-muv-relation_data}) from the fit because these contain too few objects to robustly constrain the relation. We obtain the best linear fit to the median bins via a Monte-Carlo Markov Chain (MCMC) analysis with $10^6$ steps, using the public software package \texttt{emcee} \citep{foreman-mackey13}. The best-fit mass-luminosity relation and its 1$\sigma$-range for the reference model are shown as a red line and red shaded area in Fig.~\ref{fig:m-muv-relation_data}. We proceed in the same manner for all the SED-fitting configurations detailed in section~\ref{sec:assumptions}. The best-fit parameters and their 1$\sigma$ uncertainties are summarized in Table~\ref{tab:M-Muv-Relations}.

We compare the resulting mass-luminosity relations in Fig.~\ref{fig:m-muv-relations}. The upper left panel shows the resulting mass-luminosity relations for the three different metallicities used in our analysis. The relations for the three metallicities do not differ significantly, all lying within the uncertainties in the medians of the masses. There is a trend towards systematically larger masses and larger scatter when assuming higher metallicity \citep[due to increasing mass-to-light ratios with increasing metallicity,][]{bc03}. Fig.~\ref{fig:m-muv-relations} also shows mass-luminosity relations resulting from SED-fits with different assumptions about the SFH. While the three exponential SFHs yield similar slopes as the constant SFH, only the delayed SFH results in similar masses. The exponentially rising SFH instead yields distinctly lower ($\sim-0.3$\,dex) and the exponentially declining SFH distinctly higher ($\sim0.2$\,dex) stellar masses but the exact offset depends on $M_{UV}$. The lower left panel in Fig.~\ref{fig:m-muv-relations} shows the impact of changing our assumptions for the dust attenuation in the mass-luminosity relation. We find that allowing for large dust attenuation optical depths in our sample leads to a remarkably shallower mass-luminosity relations compared to using the parameters in the reference model. Applying an SMC-like extinction law however yields masses closer to the reference model. Note that the three relations roughly result in the same masses at the bright end, meaning that dust attenuation mostly impacts the mass estimates of the faint, i.e., low-mass, galaxies.

Finally, in Fig.~\ref{fig:m-muv-relations}, we compare our $M_{\star}-M_{UV}$-relations to the most recent results in the literature, namely \citet{song16}, \citet{bhatawdekar19}, and \citet{kikuchihara20}. Our reference \texttt{BEAGLE} configuration has a similar relation to \citet{bhatawdekar19} with slightly higher masses. Both \citeauthor{song16} and \citeauthor{kikuchihara20} find steeper mass-luminosity relations. Our results depart from the literature for different SFH and dust content. We will further discuss these features in section~\ref{sec:discussion}. Note that our and the $M-M_{UV}$-relations in the literature yield similar masses at the bright end, e.g., at $M_{UV}=-19.5$. This indicates that stellar masses at the bright end are relatively insensitive ($\sim$0.5\,dex) to the assumed SED modeling parameters and that the most significant differences ($\sim$1\,dex) occur for intrinsically, faint low stellar mass galaxies.

\section{High-redshift Galaxy Stellar Mass Functions} \label{sec:GSMF}

\begin{figure}
    \centering
    \includegraphics{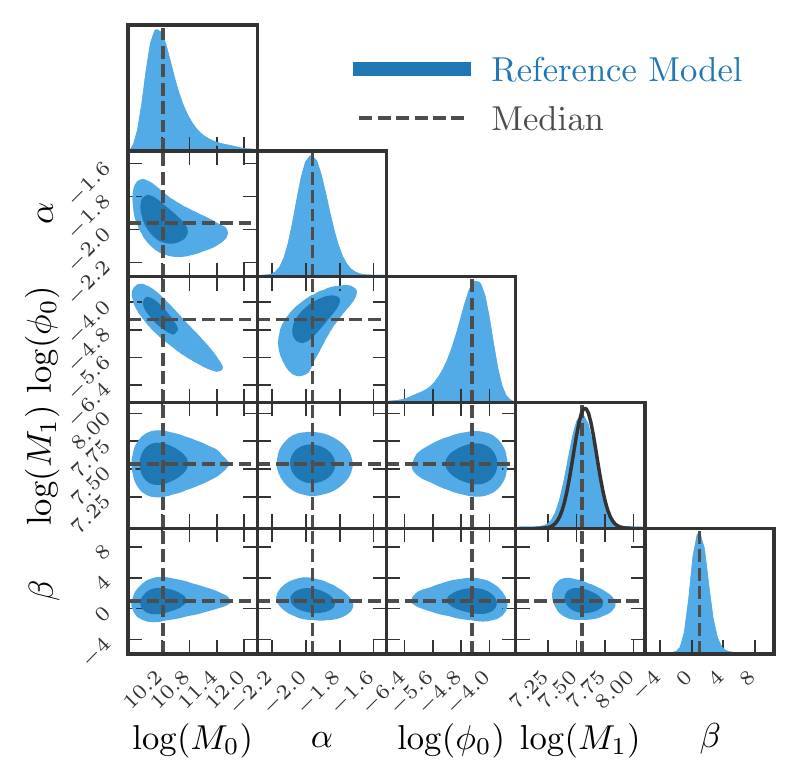}
    \caption{Posterior distributions (blue shaded areas) and median values (dashed lines) of the five modified Schechter fit parameters for the reference model (cf. Table~\ref{tab:SED-fit_parameters}). These diagrams illustrate that the fit parameters are in general, not independent, in particular the three 'classical' Schechter parameters $M_0$, $\alpha$ and $\phi_0$. The solid line in the $M_1$ panel shows the Gaussian prior applied to $M_1$.}
    \label{fig:corner_reference-model}
\end{figure}

\begin{figure*}
    \centering
    \includegraphics[width=\textwidth, keepaspectratio=True]{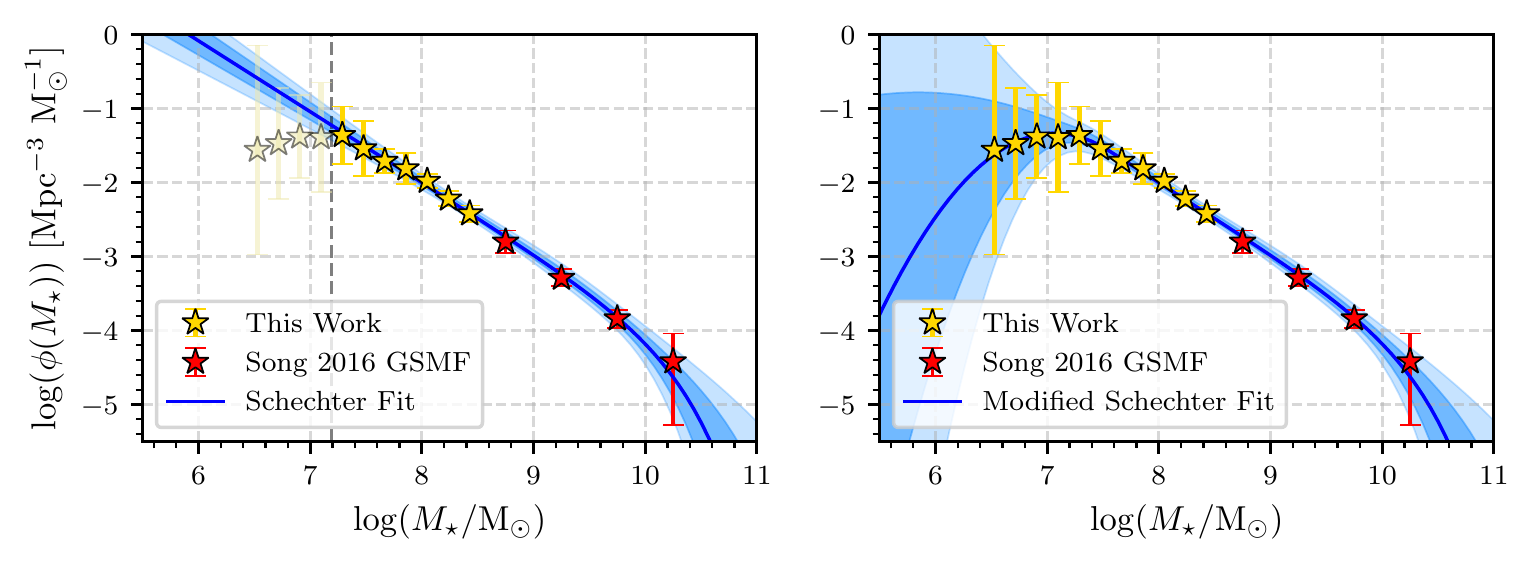}
    \caption{Final GSMFs at $z\sim6-7$ as determined using the parameters from our reference model (Table~\ref{tab:SED-fit_parameters}). Golden colored stars represent the results derived for our $z\sim6-7$ sample in the HFF and red stars represent the determination of the high-mass end of the GSMF by \citet{song16} using wide-area blank fields. We shifted the \citet{song16} results down by 0.15\,dex in order to account for the redshift-evolution between their $z\sim6$ sample and our $z\sim6-7$ sample and include them into our fitting procedure. In the \textit{left-hand} panel we fit the GSMF with a Schechter function (Eq.~\ref{eq:schechter}). Since the points on the very low-mass end clearly depart from the regular Schechter form, we restrict the fit to mass bins $\log(M_{\star}/\text{M}_{\sun})>7.2$ (delimited by the grey dotted line) which corresponds to $M_{UV}=-15$. The mass bins excluded from the Schechter fit are shown as pale golden stars. In the \textit{right-hand} panel, we show the fit to all of the GSMF bins with a modified Schechter function (Eq.~\ref{eq:modified_schechter}) which additionally includes a component for a turnover below $M_1$. The blue shaded areas in both panels indicate the $1\sigma$ and $2\sigma$ uncertainty ranges as determined from our MCMC analysis.}
    \label{fig:GSMF-data-fit}
\end{figure*}

\begin{table*}
	\caption{Best-fit Schechter (Eq.~\ref{eq:schechter}) and modified Schechter (Eq.~\ref{eq:modified_schechter}) function parameters. The best-fit parameter values are medians and their $1\sigma$-uncertainties are estimated from our MCMC analysis.}
	\label{tab:schechter-parameters}
\begin{tabular}{lcccccc}
\hline
\texttt{BEAGLE}                     &   $\log M_0$              &   $\alpha$                    &   $\log \phi_0$               &   $\log M_1$     &   $\beta$      &   $\log M_T$\\
Configuration                       &   (M$_{\sun}$)            &                               &   (Mpc$^{-3}$)                &   (M$_{\sun}$)   &                &   (M$_{\sun}$)\\
\hline
\multicolumn{7}{c}{\textit{Schechter Eq.~\eqref{eq:schechter} fit}}\\
\hline
Reference Model                     &   $10.21_{-0.26}^{+0.43}$  &   $-1.96_{-0.08}^{+0.09}$     &   $-4.48_{-0.31}^{+0.62}$     &   -                       &   -               &   -\\
$Z=0.01\,\text{Z}_{\sun}$           &   $10.21_{-0.26}^{+0.45}$  &   $-1.92_{-0.08}^{+0.08}$     &   $-4.46_{-0.31}^{+0.61}$     &   -                       &   -               &   -\\
$Z=0.5\,\text{Z}_{\sun}$            &   $10.22_{-0.26}^{+0.44}$  &   $-2.04_{-0.08}^{+0.09}$     &   $-4.54_{-0.32}^{+0.68}$     &   -                       &   -               &   -\\
$\psi(t)\propto t\exp{(-t/\tau)}$   &   $10.21_{-0.26}^{+0.44}$  &   $-1.96_{-0.08}^{+0.09}$     &   $-4.48_{-0.31}^{+0.63}$     &   -                       &   -               &   -\\
$\psi(t)\propto\exp{(t/\tau)}$      &   $10.19_{-0.26}^{+0.47}$  &   $-1.82_{-0.07}^{+0.08}$     &   $-4.38_{-0.30}^{+0.54}$     &   -                       &   -               &   -\\
$\psi(t)\propto\exp{(-t/\tau)}$     &   $10.23_{-0.26}^{+0.39}$  &   $-2.14_{-0.09}^{+0.10}$     &   $-4.60_{-0.32}^{+0.74}$     &   -                       &   -               &   -\\
Calzetti attenuation law            &   $10.37_{-0.29}^{+0.42}$  &   $-2.34_{-0.10}^{+0.11}$     &   $-5.00_{-0.34}^{+1.04}$     &   -                       &   -               &   -\\
SMC-like extinction law             &   $10.33_{-0.29}^{+0.45}$  &   $-2.11_{-0.08}^{+0.09}$     &   $-4.75_{-0.33}^{+0.82}$     &   -                       &   -               &   -\\
\hline
\multicolumn{7}{c}{\textit{Modified Schechter Eq.~\eqref{eq:modified_schechter} fit}}\\
\hline
Reference Model                     &   $10.22_{-0.27}^{+0.45}$  &   $-1.96_{-0.08}^{+0.09}$     &   $-4.49_{-0.32}^{+0.64}$     &   $7.55_{-0.10}^{+0.10}$  &   $1.00_{-0.73}^{+0.87}$  &   $7.10_{-0.56}^{+0.17}$\\
$Z=0.01\,\text{Z}_{\sun}$           &   $10.22_{-0.26}^{+0.47}$  &   $-1.92_{-0.08}^{+0.08}$     &   $-4.48_{-0.32}^{+0.62}$     &   $7.50_{-0.07}^{+0.07}$  &   $0.91_{-0.74}^{+0.80}$  &   $7.04_{-0.65}^{+0.18}$\\
$Z=0.5\,\text{Z}_{\sun}$            &   $10.23_{-0.27}^{+0.43}$  &   $-2.04_{-0.09}^{+0.09}$     &   $-4.55_{-0.32}^{+0.69}$     &   $7.67_{-0.12}^{+0.12}$  &   $1.18_{-0.78}^{+0.98}$  &   $7.25_{-0.46}^{+0.15}$\\
$\psi(t)\propto t\exp{(-t/\tau)}$   &   $10.19_{-0.26}^{+0.42}$  &   $-1.96_{-0.08}^{+0.09}$     &   $-4.45_{-0.31}^{+0.62}$     &   $7.36_{-0.30}^{+0.31}$  &   $1.42_{-1.14}^{+3.11}$  &   $7.01_{-0.35}^{+0.19}$\\
$\psi(t)\propto\exp{(t/\tau)}$      &   $10.19_{-0.26}^{+0.47}$  &   $-1.82_{-0.07}^{+0.08}$     &   $-4.38_{-0.30}^{+0.54}$     &   $7.29_{-0.12}^{+0.12}$  &   $0.71_{-0.83}^{+0.99}$  &   $6.85_{-0.75}^{+0.18}$\\
$\psi(t)\propto\exp{(-t/\tau)}$     &   $10.22_{-0.26}^{+0.40}$  &   $-2.14_{-0.09}^{+0.10}$     &   $-4.59_{-0.32}^{+0.74}$     &   $7.71_{-0.26}^{+0.27}$  &   $1.79_{-1.36}^{+3.39}$  &   $7.38_{-0.32}^{+0.17}$\\
Calzetti attenuation law            &   $10.37_{-0.29}^{+0.43}$  &   $-2.34_{-0.10}^{+0.11}$     &   $-4.97_{-0.35}^{+1.04}$     &   $8.01_{-0.18}^{+0.21}$  &   $1.37_{-8.47}^{+10.16}$ &   $7.86_{-0.31}^{+0.14}$\\
SMC-like extinction law             &   $10.33_{-0.29}^{+0.45}$  &   $-2.10_{-0.08}^{+0.09}$     &   $-4.74_{-0.33}^{+0.81}$     &   $7.77_{-0.17}^{+0.18}$  &   $1.56_{-2.04}^{+3.85}$  &   $7.49_{-0.44}^{+0.14}$\\
\hline
\end{tabular}
\end{table*}

\begin{figure*}
    \centering
    \includegraphics[width=\textwidth, keepaspectratio=True]{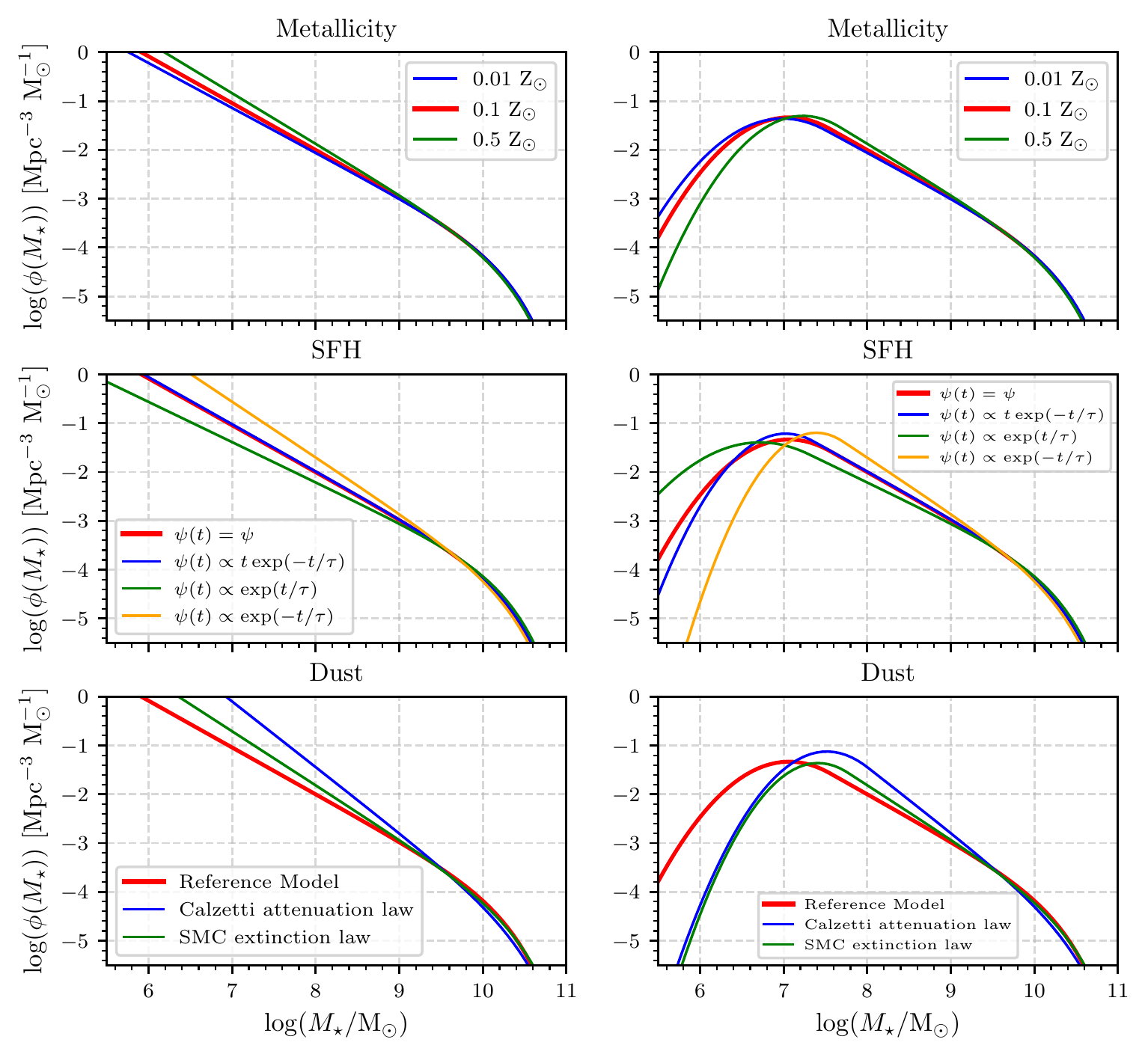}
    \caption{Best-fit GSMFs for each \texttt{BEAGLE} configuration explored in this work (see section~\ref{sec:assumptions}). The data are fitted with a Schechter function (Eq.~\ref{eq:schechter}) in the \textit{left-hand} panels and with a modified Schechter function (Eq.~\ref{eq:modified_schechter}) in the \textit{right-hand} panels. The GSMF for the reference model is plotted in red in each panel. The label at the top of each panel indicates which parameter is investigated and the legends indicate the parameter values.}
    \label{fig:GSMFs}
\end{figure*}

We now use the $M_{\star}-M_{UV}$-relations derived in section~\ref{sec:M-Luv-relations} to convert the observed rest-frame UV luminosity function to galaxy stellar mass functions (GSMFs). This approach is commonly used in the study of high-redshift GSMFs because selection and completeness effects are well-studied for UV luminosity functions \citep[e.g.][]{gonzales11,song16,kikuchihara20}.

To accomplish this transformation, we take the luminosity function from \citet{atek18}, which was derived from the same HFF $z\sim6-7$ sample that we use in this study, and transform the luminosity bins into stellar mass bins using the median $M_{\star}-M_{UV}$-relations. The GSMF uncertainties in each of the luminosity/stellar mass bins include contributions from the uncertainties in the survey volume and cosmic variance \citep[see][for details]{atek18}. In addition, we account for uncertainties in the determination of stellar mass with \texttt{BEAGLE} by making $10^4$ random realizations drawn from each galaxy's stellar mass posterior distribution and re-binning in stellar mass at each iteration. Doing this means that galaxies, given the uncertainties, can shift between mass bins from iteration to iteration thus providing a robust estimate of the uncertainties in the number of galaxies in each bin and on the final median stellar mass of any particular bin.

We show the resulting GSMF bins for the reference model in Fig.~\ref{fig:GSMF-data-fit}. All of our final GSMF determinations and their uncertainties can be found in Tables~\ref{tab:GSMF_points} and \ref{tab:GSMF_errors} in appendix~\ref{app:GSMF-points}. We observe a possible turnover of the GSMF for masses below $\log(M_{\star}/\text{M}_{\sun})\lesssim7$ which corresponds to the turnover in the \citet{atek18} UV luminosity functions. However, a steepening or no change at all in the slope of the GSMF is less likely but also consistent within the relatively large uncertainties. A low-mass turnover does not appear in any high-redshift GSMFs in the literature. It is however present in the HFF high-redshift UV luminosity functions \citep{bouwens17a,atek18}. We combine our results with GSMF bins derived by \citet{song16}, who combined results from CANDELS, GOODS and HUDF fields to increase the area analyzed and were able to constrain the high-mass end of the GSMF \citep[see also][for a similar analysis]{kikuchihara20}. Following \citet{bouwens17a} and \citet{atek18}, we shift their $z\sim6$~GSMFs down by 0.15\,dex in order to account for the redshift-evolution of the GSMF from their sample to our $z\sim6-7$ sample.

To fit the observed GSMF bins we adopt a standard Schechter function \citep{schechter76}. The general form of the Schechter function is,

\begin{equation} \label{eq:schechter}
    \phi(M_{\star})=\ln({10})\phi_0\left(\frac{M_{\star}}{M_0}\right)^{\alpha+1}e^{-\frac{M_{\star}}{M_0}}
\end{equation}

\noindent
where $M_0$ corresponds to the characteristic stellar mass where the power-law turns into an exponential, $\alpha$ is the low-mass end slope and $\phi_0$ the normalization. Since the observed GSMF bins depart from the classical Schechter form at $\log(M_{\star}/\text{M}_{\sun})<7.2$ (grey dashed line in Fig.~\ref{fig:GSMF-data-fit}), we also fit the GSMFs with a modified Schechter function \citep{bouwens17a,atek18} which accounts for a potential turnover at low masses. In the modified Schechter function we multiply the expression in Eq.~\eqref{eq:schechter} with a turnover term,

\begin{equation} \label{eq:turnover-term}
    10^{-\beta\log\left(\frac{M_{\star}}{M_1}\right)^2}
\end{equation}

\noindent
for stellar masses lower than $M_1$. This results in a modified Schechter function as,

\begin{equation} \label{eq:modified_schechter}
    \phi(M_{\star})=\begin{cases} \ln({10})\phi_0\left(\frac{M_{\star}}{M_0}\right)^{\alpha+1}e^{-\frac{M_{\star}}{M_0}} & , M_{\star}>M_1\\
    \ln({10})\phi_0\left(\frac{M_{\star}}{M_0}\right)^{\alpha+1}e^{-\frac{M_{\star}}{M_0}}10^{-\beta\log\left(\frac{M_{\star}}{M_1}\right)^2} & , M_{\star}\leq M_1
    \end{cases}
\end{equation}

\noindent
where the curvature parameter $\beta$ allows for a downward turnover of the GSMF if $\beta>0$ and for an upward turnover if $\beta<0$. We define the turnover mass $M_T$ as the stellar mass corresponding to the maximum in the GSMF in the case of $\beta>0$, i.e., where $(\text{d}\phi/\text{d}M_{\star})_{M_{\star}=M_T}=0$. We derive the best-fit Schechter and modified Schechter parameters as the median of the joint posterior distribution of 20 MCMC chains with $5\times10^5$ steps (after removing the first $\sim$20000 steps as burn-in phase). The likelihood used in this inference can be expressed as,

\begin{equation} \label{eq:GSMF-likelihood}
    \mathcal{L}=\prod_{i}\frac{1}{\phi_i}\frac{1}{\sqrt{2\pi}\sigma_i}e^{\frac{(\ln{\phi_i}-\ln{\phi(M_{\star,i}}))^2}{2\sigma_i^2}}
\end{equation}

\noindent
where $\phi_i$ and $\sigma_i$ are the observed GSMF bins and their uncertainty respectively and $\phi(M_{\star,i})$ is the GSMF model function evaluated at each stellar mass bin. Note that we only consider $\log(M_{\star}/\text{M}_{\sun})>7.2$ bins when fitting the Schechter function \eqref{eq:schechter} since the mass bins below that (pale golden stars in the left-hand panel of Fig.~\ref{fig:GSMF-data-fit}) clearly depart from the 'classical' Schechter form and therefore cannot be fit with Eq.~\eqref{eq:schechter}. For the reference model, fitting all mass bins with Eq.~\eqref{eq:schechter} however only marginally impacts the resulting parameters (cf. Fig.~\ref{fig:GSMF_classical_schechter_full} in appendix~\ref{app:GSMF-points}). The modified Schechter function, Eq.~\eqref{eq:modified_schechter}, on the other hand is fit to all mass bins. We assume flat priors for the fit parameters, $M_0$, $\alpha$, $\phi_0$ and $\beta$. \citet{atek18} found that the best fit to the UV luminosity functions introduces the turnover term for magnitude bins brighter than $M_{UV}=-16$. To fit the modified Schechter GSMFs, we therefore use a Gaussian prior on $M_1$ with the stellar mass that corresponds to $M_{UV}=-16$ and the uncertainties in the $M_{\star}-M_{UV}$-relation respectively as the peak and the standard deviation of the distribution. The parameter spaces are limited to $8<\log(M_0/\text{M}_{\sun})<12$, $-3<\alpha<-1$, $-10<\log(\phi_0/\text{Mpc}^{-3})<0$, $6<\log(M_1/\text{M}_{\sun})<9$ and $-20<\beta<20$.

The best-fit parameters and their 1$\sigma$ uncertainties are listed in Table~\ref{tab:schechter-parameters} for all the different SED-fitting runs and we show the posterior distributions of the four modified Schechter parameters for the reference model in Fig.~\ref{fig:corner_reference-model}. Note that the fit-parameters are not independent in general. In particular, there is a degeneracy between the high-mass exponential cutoff $M_0$ and the low-mass end slope, $\alpha$. It therefore appears that knowledge of both the high- and the low-mass ends are required to constrain the overall shape of the GSMF. We show best-fit Schechter and modified Schechter functions for the reference model in Fig.~\ref{fig:GSMF-data-fit}.

The best-fit Schechter and modified Schechter functions for each \texttt{BEAGLE} configuration in Table~\ref{tab:schechter-parameters} are plotted in Fig.~\ref{fig:GSMFs} and we show the corresponding posterior distributions in Fig.~\ref{fig:corner_plots} in appendix~\ref{app:GSMF-points}. The resulting GSMFs do not depend significantly on metallicity within their respective uncertainties. We observe a slight steepening of the low-mass end slope with increasing metallicity, from $\alpha\simeq-1.92_{-0.08}^{+0.08}$ for $0.01\,\text{Z}_{\sun}$ to $\alpha\simeq-2.04_{-0.09}^{+0.09}$ for $0.5\,\text{Z}_{\sun}$, resulting in a insignificantly larger turnover mass $\log(M_T/\text{M}_{\sun})\simeq7.25_{-0.46}^{+0.15}$ and curvature $\beta\simeq1.18_{-0.78}^{+0.98}$ for the highest metallicity case. The effect of changing our assumptions on the form of the SFH on the GSMF is much more severe: While assuming a delayed SFH results in the same GSMF as a constant SFH, the exponentially rising and declining SHFs result in significantly shallower ($\alpha\simeq-1.82_{-0.07}^{+0.08}$) and steeper ($\alpha\simeq-2.14_{-0.09}^{+0.10}$) GSMFs, respectively. Finally, allowing for more dust attenuation in the ensemble of galaxies results in a much steeper GSMF, $\alpha\simeq-2.34_{-0.10}^{+0.11}$, than the reference model. Applying an SMC dust extinction law instead of the Calzetti law however results in a GSMF closer to the reference model in the low-mass end slope, $\alpha\simeq-2.10_{-0.08}^{+0.09}$. Larger dust attenuation insignificantly affects the high-mass end with an exponential cut-off at $\log(M_0/\text{M}_{\sun})\simeq10.37_{-0.29}^{+0.43}$ instead of at $\log(M_0/\text{M}_{\sun})\simeq10.22_{-0.27}^{+0.45}$ in the reference model. We further note a slight correlation between the low-mass end slope, $\alpha$, and the turnover curvature, $\beta$, in general: A steeper slope $\alpha$ results in a larger turnover mass $M_T$ and a slightly higher curvature $\beta$. The uncertainties in $\beta$ become particularly large for the steepest $\alpha$. This is due to the $M_{\star}$-bins lying closer together in these cases because the $M_{\star}-M_{UV}$-relation is shallower. The large uncertainties on the lowest-mass bins therefore allow for a much wider range of acceptable values in the $\beta$-parameter space (cf. lower panel of Fig.~\ref{fig:corner_plots}). While our modified Schechter parametrization of the GSMF also allows for an upward turnover, i.e., $\beta<0$, this is ruled out by our MCMC analysis at greater than 1$\sigma$-level for all models apart from the exponentially rising SFH case (cf. Fig.~\ref{fig:GSMF-rising_SFH}) and the two models that allow higher values of the dust attenuation although this is a result of the effects already discussed above in these two models.

\section{Discussion}\label{sec:discussion}

We presented a determination of the stellar masses of a sample of $z\sim6-7$ galaxies observed through the gravitational magnification of the six HFF clusters and computed their GSMFs from UV luminosity functions \citep{atek18}. We found relatively shallow $M_{\star}-M_{UV}$-slopes $\sim0.4$, consistent with constant mass-to-light ratios. Our GSMFs have relatively steep low-mass end slopes and relatively low exponential cut-offs with $\log(M_0/\text{M}_{\sun})\simeq10.22_{-0.27}^{+0.45}$, $\alpha\simeq-1.96_{-0.08}^{+0.09}$ and $\log(\phi_0/\text{Mpc}^{-3})\simeq-4.49_{-0.32}^{+0.64}$ for the reference SED-fitting model. We also observe a turnover at the very low-mass end of the GSMF at $\log(M_T/\text{M}_{\sun})\simeq7.10_{-0.56}^{+0.17}$ with a downward curvature parameter $\beta\simeq1.00_{-0.73}^{+0.87}$. In this section, we discuss these results with regard to the IR (rest-frame optical) photometry, gravitational lensing uncertainties and the impact of SED-fitting prescriptions (sections~\ref{sec:phot-discussion}, \ref{sec:gravlens-discussion} and \ref{sec:degeneracies}). Finally, we compare our results with those from the literature and we reflect briefly on the nature of the turnover in the GSMF (sections~\ref{sec:literature} and \ref{sec:theory}).

\subsection{Photometry}\label{sec:phot-discussion}

The rest-frame UV emission in star forming galaxies is dominated by massive O and B stars. Since these stars have very short lifetimes, they critically depend on the recent star-forming activity. One needs to observe wavelengths beyond the 4000\,\AA~break in the rest-frame optical which preferentially probe the photospheric emission from older stellar populations. 
In this study, we use deep \textit{Spitzer}/IRAC data which sample the emission red-ward of $\lambda\sim4000$\,\AA~of our galaxies. While these rest-frame \textit{optical} photometric bands are crucial in estimating stellar masses at $z>6$ \citep[for a complete review see e.g.,][]{dunlop13}, they are affected by the assumed SFH as we discuss in Section~\ref{sec:SFH-impact}.

Obtaining reliable IRAC photometry is challenging because of two main problems: The limited depth of the \textit{Spitzer} observations, $\lesssim26$\,mag, (Table~\ref{tab:field-depths}) and blending with the light profiles of foreground objects. The latter problem is severe due to the IRAC observations having $10\times$ lower resolution than the HST observations used to select high-redshift targets. Since our $z\sim6-7$ galaxies are mostly unresolved, even in HST detection stacks, with angular sizes $\lesssim1\arcsec$ the pixel scale of the HFF IRAC mosaics of $0.6\arcsec$ and the IRAC PSFs with $\sim2\arcsec$ FWHM make them blend into the much higher surface brightness light profiles of foreground galaxies. This effect is moreover amplified by the very nature of our observations: Since we observe the high-redshift galaxies \textit{through} the dense cores of SL clusters we are looking at exceptionally crowded fields, dominated by ICL and cluster galaxies. While we correct for unreliable IRAC photometric measurements by an empirical correction factor (see section~\ref{sec:IRAC-photometry_correction}), this correction is an average over a relatively small (only 39) subsample of our full sample, 303 (13\%). There is a significant scatter in this correction even among sources with reliable IRAC photometry, of about $S_{\delta}\sim0.2$\,dex (cf. Fig.~\ref{fig:IRAC-photometry_correction}). The $z\sim6-7$ GSMF is sensitive to fluctuations in the overall $M_{\star}$ distribution of that order of magnitude (cf. e.g. section~\ref{sec:SFH-impact}). Furthermore, we expect a slight bias towards overestimating the lower stellar masses due to selection effects in the sample of galaxies for which we derive the mean correction factor. This bias is difficult to quantify because we can not constrain the actual stellar mass of these objects at the low-mass end without IRAC photometry. We however expect its impact on our results to be relatively small compared to the large uncertainties in the estimated values of the lensing magnification (cf. appendix~\ref{app:irac-correction_bias} for further discussion).

The next generation of high angular resolution facilities such as the \textit{James Webb Space Telescope} \citep[JWST;][]{gardner06} will be able to observe the rest-frame optical continuum and nebular emission lines of these $z\sim6-7$ galaxies with angular resolutions and depths comparable to the current HST observations of distant lensed galaxies. Recent simulations have shown that rest-frame optical photometry with the \textit{Near Infrared Camera} \citep[NIRCam;][]{rieke05} aboard the JWST will be able to constrain the stellar mass of galaxies out to $z\sim9$ to within 0.2\,dex \citep{bisigello17,kemp19,kauffmann20}. This will allow us to robustly constrain the low-mass end of the galaxy mass stellar function.

\subsection{Gravitational lensing uncertainties}\label{sec:gravlens-discussion}

\begin{figure}
    \centering
    \includegraphics[width=0.5\textwidth, keepaspectratio=True]{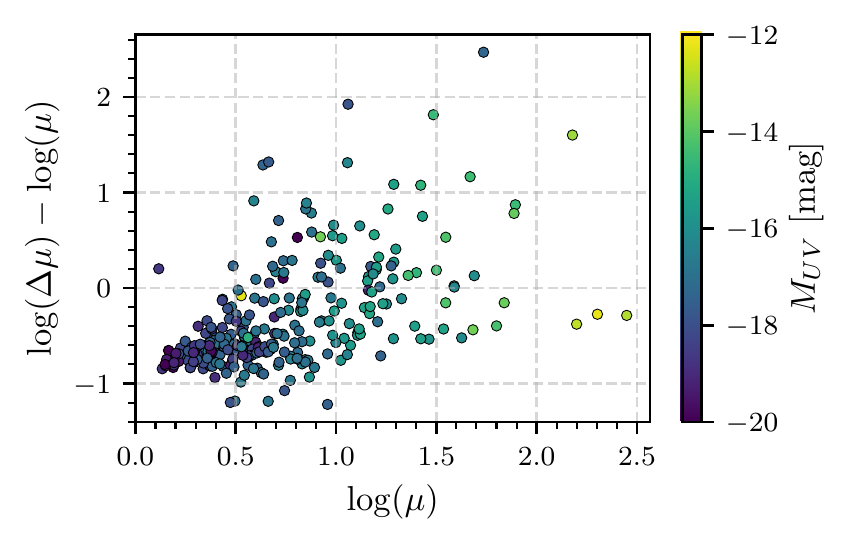}
    \caption{Relative magnification uncertainties $\frac{\Delta\mu}{\mu}$ as a function of magnification $\mu$, color-coded with UV luminosity $M_{UV}$ in AB magnitudes. There is a general trend of large relative magnification errors for large $\mu$ due mostly to SL modelling systematics between models. A few sources in the very high magnification regime \textit{lower right quadrant} are of particular interest because the major contribution to their magnification comes from the SL cluster BCGs}
    \label{fig:mu-errors}
\end{figure}

Perhaps the most significant problem inherent in studies of strongly lensed galaxies are the very large differences in the strong lensing models. In our study, we consider these differences as if they were random uncertainties and this is one of our central assumptions. This uncertainty has a significant impact on our conclusions about the robustness of the GSMF, especially at its low-mass (intrinsically faint) end. To illustrate these details, we show in Figure~\ref{fig:mu-errors} the \textit{relative} magnification uncertainties, i.e., $\frac{\Delta\mu}{\mu}$, as a function of the median magnification $\mu$ for each object. As expected, the more strongly magnified galaxies, with $\mu>10$ that have the faintest UV luminosities $M_{UV}\gtrsim-16$ (Fig.~\ref{fig:mu-errors}), also tend to have the larger uncertainties in their magnifications. This sensitivity in galaxies with high magnifications is due to their generally close proximity to the caustics. The positions and shapes of the caustics and their concomitant high magnification are very sensitive to the positions of the cluster galaxies and the assumed total mass and mass profiles of both the cluster dark matter (DM) halo and the cluster galaxies. Large numbers of multiply imaged galaxies with accurate spectroscopic redshifts are therefore required to robustly constrain the shapes and positions of the caustics and the resulting (high) magnifications from the strong lensing models. The quality and quantity of the multiple images used for constraining the cluster SL model can bias magnifications of $\mu>2$ by up to 60\,\% within the same SL modelling technique alone \citep{acebron18}. The caustics in turn tend to have radically different shapes and positions between different SL models, the number of multiple image redshifts available for the SL model, assumptions on cluster DM distributions, etc. This adds large systematic uncertainties to the magnifications which are inherent to the modelling techniques and are, as yet, not very well understood \citep{meneghetti17,acebron17}. Since we compute the magnification uncertainty, $\Delta\mu$, as the scatter between the different SL models used in this study (see section~\ref{sec:SL}), we find that the lensing uncertainties are dominated by the systematic differences in the lensing models. There are a few sources on the very high magnification end which have distributions in magnification that are of the same order as the median magnification (lower right quadrant in Fig.~\ref{fig:mu-errors}). Visual inspection revealed that these sources are lensed by massive brightest cluster galaxies (BCGs) in the dense centers of the clusters. Significant improvements of their magnifications could therefore come from dynamic modelling of the BCGs and the mass potential within these clusters using Integral Field spectroscopy \citep[IFS; e.g.,][]{chirivi20}.

While gravitational lensing by massive clusters and their galaxies allows us to \textit{detect} the faintest $z\gtrsim6$ galaxies, it introduces large systematic uncertainties in de-magnified magnitudes and therefore the luminosities and masses. Improvements in SL modelling techniques \citep[such as proposed by e.g.,][]{niemiec20,chirivi20,yang20} and large samples of spectroscopic redshifts of lensed galaxies, especially those with multiple images, are needed to reduce such uncertainties and differences in the models.
 
\subsection{Impact of SED-fitting assumptions}\label{sec:degeneracies}

We describe the impact of SED-fitting assumptions and constraints on stellar mass and the resulting $M_{\star}-M_{UV}$-relation in section~\ref{sec:M-Luv-relations} (Fig.~\ref{fig:m-muv-relations} and Table~\ref{tab:M-Muv-Relations}) and on the resulting GSMF in section~\ref{sec:GSMF} (Fig.~\ref{fig:GSMFs} and Table~\ref{tab:schechter-parameters}). While we found that the stellar mass (and thus the GSMF) is not strongly dependent on metallicity, we did find significant differences depending on the underlying assumption as to the functional form of the SFH, the range of attenuation optical depths which is allowed in the modeling, and the nature of the attenuation law. We individually address these degeneracies in detail in the following sections.

\subsubsection{SFH} \label{sec:SFH-impact}

\begin{figure}
    \centering
    \includegraphics[width=\columnwidth, keepaspectratio=True]{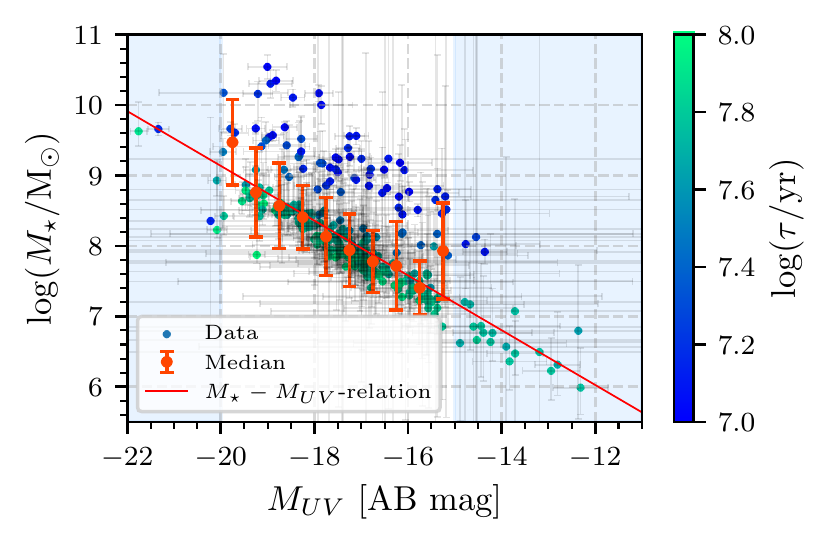}
    \caption{$M_{\star}-M_{UV}$ distribution for a delayed SFH, $\psi(t)\propto t\exp{(-t/\tau)}$. The values are color-coded by the SFR $e$-folding time, $\tau$, as indicated by the color bar on the right. The red line shows the best-fit $M_{\star}-M_{UV}$-relation to median-binned data. With this SFH, the galaxy sample separates into two distinct stellar mass populations which illustrates the degeneracy between $M_{\star}$ and $\tau$: Galaxies with low mass close to their SFR peak at $t_{\text{age}}=\tau$ (green) and galaxies with high mass which are not currently close to the peak in their SFR (blue). This scatter does not appear for a constant SFH (cf. Fig.~\ref{fig:m-muv-relation_data}). The blue shaded area is the same as in Fig.~\ref{fig:m-muv-relation_data}.}
    \label{fig:tau-degeneracy}
\end{figure}

Assuming an exponentially rising SFH, $\psi(t)\propto\exp{(t/\tau)}$, results in stellar masses lower by $\sim0.3$\,dex than a constant SFH (and thus in a shallower GSMF). This is due to the radically different functional form of the exponentially rising SFH compared that used in our reference model. Since the SFR is exponentially \textit{rising} over time, the galaxies are generally fitted with larger current SFRs and thus much larger EWs in optical nebular emission lines such as the \ion{O}{iii} and H\,$\mathrm{\beta}$ lines. Large EW nebular emission then contributes a large fraction of the observed flux in the rest-frame optical bands. The rest-frame optical bands play a significant role in constraining both the ages and masses of galaxies. Very large SFRs, which result in large EWs of the optical emission lines, then result in lower estimated ages and hence stellar masses for galaxies at these redshifts. Note that the $\sim0.3$\,dex difference in $M_{\star}$ that we find agrees with the predictions by hydrodynamic simulations of the effect of changing the functional form of the SFH to an exponentially rising form \citep[$\lesssim0.3$\,dex;][]{finlator07}. Similarly, assuming an exponentially \textit{declining} SFH $\psi(t)\propto\exp{(-t/\tau)}$ has the opposite effect, the current SFR estimates are generally lower and thus this assumption results in overall larger stellar mass estimates than our reference model. The nebular emission contributes less to the flux in the rest-frame optical photometry and the thus higher estimated optical continuum luminosities yield larger stellar masses.

The delayed SFH, $\psi(t)\propto t\exp{(-t/\tau)}$, on the other hand yields the same $M_{\star}-M_{UV}$-relation as the constant SFH and thus the same GSMF. This is however mainly an effect of the median-binning we use to determine the $M_{\star}-M_{UV}$-relation (section~\ref{sec:M-Luv-relations}). The individual stellar masses scatter significantly (Fig.~\ref{fig:tau-degeneracy}) compared to the constant SFH case (Fig.~\ref{fig:m-muv-relation_data}). The galaxies separate into two populations: Galaxies with low mass and large SFR $e$-folding time, $\tau$, and galaxies with large mass and low $\tau$. Since the delayed SFH has a maximum SFR at $t=\tau$, \texttt{BEAGLE} can fit the same IRAC photometry either with a galaxy close to its SFR peak, i.e., with a large $\tau$, resulting in smaller stellar mass \textit{or} with a galaxy long after its SFR peak, i.e., with a small $\tau$, which results in larger stellar masses for the same reason discussed above for the exponential SFHs. Applying any of the three exponential SFHs therefore creates a well known degeneracy between $M_{\star}$ and $\tau$ which cannot be resolved with the available data, effectively making $\tau$ a 'nuisance' parameter in this type of analysis.

While constant or exponentially decreasing SFHs were previously assumed \citep[e.g.,][]{eyles07,stark09,gonzales11,grazian15,kikuchihara20}, results from hydrodynamical simulations predict exponentially increasing SFRs for high-redshift galaxies \citep[e.g.,][]{finlator11}. Observations of very high EW optical emission lines of [\ion{O}{iii}] and H$\mathrm{\beta}$, up to rest-frame EW\,$\sim2000$\,\AA, \citep[e.g.,][]{atek11,atek14b,reddy18b} and of IRAC excesses in high-redshift galaxies \citep[which have been attributed to strong contributions from rest-frame optical emission lines;][]{smit14,smit15,debarros19,endsley21} further support young galaxies with relatively high rates of recent star formation, which could be due to a SFH with an increasing SFR as a function of time. Alternatively, given their compact sizes, the star formation in young high-redshift galaxies might also be episodic with several relatively strong bursts occurring over relatively short duty cycles \citep[e.g.,][]{stark09}. Such episodic star formation is difficult to parametrize correctly given the relative crudeness of the data available for galaxies at these redshifts. There are however examples of models that combine a continuous SFH with an ongoing recent starburst \citep[e.g.][]{endsley21}. Other recent studies  infer the SFH from SED-fitting which is however heavily degenerate with stellar mass, age, metallicity and dust attenuation \citep{duncan14,song16,bhatawdekar19}. While \citet{grazian15} found no dependence of $M_{\star}$ on SFH in an experiment similar to ours for $z\sim3.5-4.5$ galaxies, their SED-fitting analysis did not include nebular emission and would therefore not have been sensitive to some of the degeneracies discussed above.

Progress on determining the SFHs of distant galaxies will also come with the launch of the JWST. Observations with the \textit{Near Infrared Spectrograph} \citep[NIRSpec;][]{bagnasco07,birkmann16} will enable us to resolve and accurately characterize the optical nebular emission lines at $z\sim6-7$ \citep{chevallard19,maseda19}. Observations with the \textit{Mid-Infrared Instrument} \citep[MIRI;][]{rieke15,wright15} will be the first instrument able to detect the rest-frame NIR continuum emission of these galaxies thereby enabling the most robust and most SFH-independent estimate possible of $M_{\star}$ \citep[e.g.,][]{dunlop13}.

\subsubsection{Dust attenuation} \label{sec:dust-impact}

\begin{figure}
    \centering
    \includegraphics[width=0.5\textwidth, keepaspectratio=True]{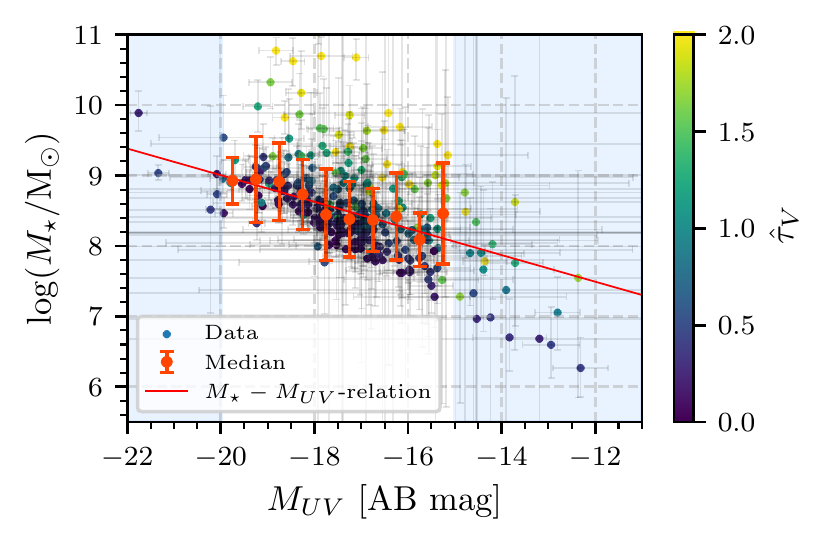}
    \includegraphics[width=0.5\textwidth, keepaspectratio=True]{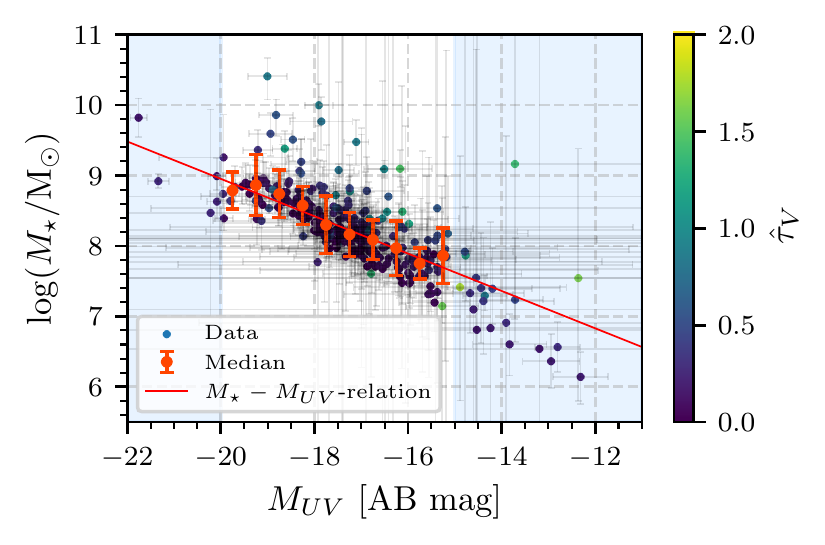}
    
    \caption{$M_{\star}-M_{UV}$ distribution for a Calzetti dust attenuation law (\textit{upper panel}) and for an SMC-like law (\textit{lower panel}) when allowing the effective V-band optical depth to vary over a wide range, $\hat{\tau}_V\in[0, 3]$. Each value is color-coded with $\hat{\tau}_V$. The red line shows the best-fit $M_{\star}-M_{UV}$-relation to median-binned data. We observe a degeneracy between $\hat{\tau}_V$ and $M_{\star}$, in particular for the Calzetti law case. For the steeper SMC-like law this degeneracy is less prominent as the galaxies are fit with much less dust attenuation, $\hat{\tau}_V\lesssim1.5$. The blue shaded area is the same as in Fig.~\ref{fig:m-muv-relation_data}.}
    \label{fig:dust-degeneracy}
\end{figure}

The other significant impact of SED-fitting parameters on the $z\sim6-7$ GSMFs comes from allowing the best fits to the SEDs of individual high-redshift galaxies to have a higher dust attenuation optical depth (Fig.~\ref{fig:m-muv-relations}). Allowing a wide range of effective \textit{V}-band optical depth, up to $\hat{\tau}_V\leq3$, results in significantly shallower $M_{\star}-M_{UV}$-slopes and thus in turn in steeper low-mass end GSMF slopes (Fig.~\ref{fig:m-muv-relations}) than in the case when the range allowed in $\hat{\tau}_V$ is more limited, i.e., $\hat{\tau}_V\leq0.2$ (section~\ref{sec:M-Luv-relations} and Fig.~\ref{fig:GSMFs}). This impact is due to the well-known degeneracy between dust attenuation and stellar mass (cf. Fig.~\ref{fig:dust-degeneracy}). This degeneracy is the reason for the considerable scatter in $M_{\star}$ in all luminosity bins and results in very large uncertainties in the $M_{\star}-M_{UV}$-relation. We however also observe that dust attenuation mostly affects the median stellar masses of the faint $M_{UV}$-bins, i.e., the bins with the highest fractions of galaxies without reliable IRAC photometry. We expected this behavior since these galaxies' stellar mass is constrained only by rest-frame UV photometry which is more strongly affected by amount of dust attenuation compared the rest-frame optical photometry \citep[the typical factor between $A_V$ and $A_{UV}$ is $\sim4$][]{calzetti00}.

As can be seen in Fig.~\ref{fig:dust-degeneracy}, this degeneracy seems to be reduced by using a steeper SMC-like dust attenuation law \citep[such as observed in e.g.,][]{reddy15,reddy18a}. Fitting the SEDs with $\hat{\tau}_V\in[0, 3]$ then results in $M_{\star}-M_{UV}$-slopes and a GSMF closer to that obtained by assuming a Calzetti attenuation law and only allowing $\hat{\tau}_V\in[0, 0.2]$ (Fig.~\ref{fig:m-muv-relations} and \ref{fig:GSMFs}). It also results in less scatter in $M_{\star}$ than the 'dusty' Calzetti law case. The $z\sim6-7$ GSMF is therefore not only highly sensitive to the range of dust attenuation allowed but also to the shape of the dust attenuation law.

In order to more robustly constrain stellar masses at $z\sim6-7$ and to limit the impact of of the details of the range of the amount and forms of the dust attenuation allowed on the GSMF, one therefore needs to consider the latest observational results. Considering outside information about the dust attenuation observed in the high-redshift galaxy population effectively places another prior on the fitting parameters. High-redshift galaxies are found to have very blue UV-continuum slopes, $\beta_{UV}\lesssim-2$, which indicates very low dust attenuations \citep{bouwens16}. This rules out our $\hat{\tau}_V\in[0,3]$ model and favors the $\hat{\tau}_V\in[0,0.2]$ model as the more realistic prior in our modeling. Note that recent observations of dust continuum emission (and [\ion{O}{iii}] and [\ion{C}{ii}] emission lines) from high-redshift galaxies with the \textit{Atacama Large Millimeter/Sub-millimeter Array} (ALMA) have discovered surprisingly high dust masses in a few high-redshift galaxies \citep{laporte19,tamura19}. This does not change our conclusions that the overall dust attenuation must be low for two salient reasons. The first is that high-redshift galaxies have very blue UV slopes which can only be explained by young ages and low dust attenuations \citep{tamura19,bakx20}. The other is that the brightest dust continuum emission regions are often not spatially coincident with the UV-bright regions of high-redshift sources. Moreover, many galaxies at these redshifts are completely undetected in the thermal IR continuum \citep[e.g.,][]{capak15}. Rest-frame far-infrared (FIR) detections with ALMA slightly favor a steep SMC-like extinction law \citep{capak15,bouwens16}. While some rest-frame UV observations of low redshift star-forming analogs support a Calzetti-like attenuation law \citep{scoville15} with a steep dependence at long wavelengths \citep[i.e., $\lambda\geq2500$\,\AA][]{reddy15}, the most recent observations strongly support an even steeper dust attenuation law than the SMC extinction law for high-redshift galaxies \citep{reddy18a}. For these reasons, there is no good reason to presuppose that the UV detected galaxies and UV bright regions embedded in what must be larger structures when the dust continuum and UV emission are not co-spatial have high attenuation optical depths such as $\hat{\tau}_V\ga$ a few tenths of a magnitude.

Additional multi-band rest-frame FIR observations with ALMA to deeper flux levels will be required to place stronger constraints on dust content and attenuation in SED-fitting stellar masses at high-redshift. Rest-frame NIR observations with JWST/MIRI will be crucial in breaking the degeneracy with dust attenuation \citep{kemp19} and stellar age since NIR wavelengths are the least affected by dust attenuation as well as having mass-to-light ratios that are less sensitive to stellar age. These additional data will enable us to minimize the impact of various degeneracies in SED modeling and ultimately to make robust estimates of $M_{\star}$ and thus the GSMF at these redshifts.

\subsection{Comparison to the literature}\label{sec:literature}

\begin{figure}
    \centering
    \includegraphics[width=0.5\textwidth, keepaspectratio=True]{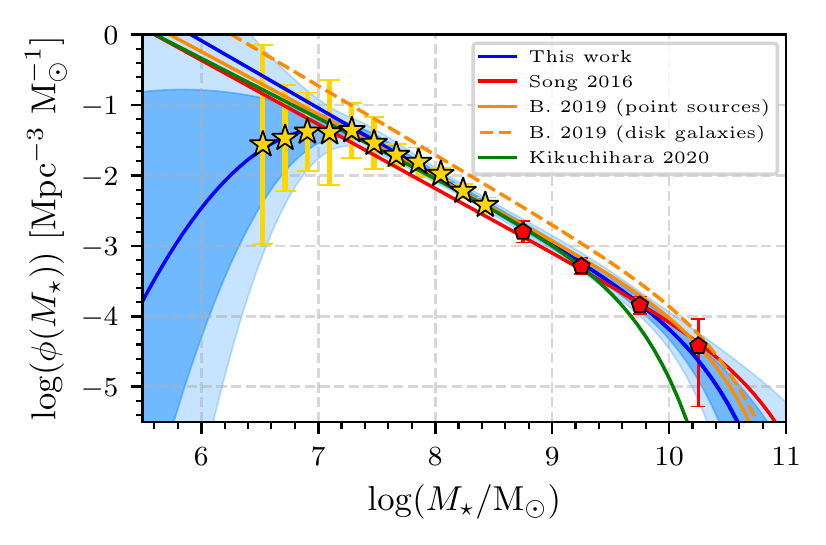}
    \caption{The final $z\sim6-7$ GSMF (for the reference \texttt{BEAGLE} configuration, blue curve) compared to GSMFs published in the recent literature. Yellow stars represent our reference model low-mass GSMF points and red pentagons the $z\sim6$ GSMF points observed in wide-area blank fields by \citet{song16} shifted down by 0.15\,dex (see text for details). The blue curves show both our best-fit Schechter and modified Schechter functions while GSMFs by \citet{song16}, \citet{bhatawdekar19} and \citet{kikuchihara20} are shown in red, orange and green respectively (see legend at the top right). The red and purple curves are shifted down by 0.15\,dex in order to account for the difference in mean redshift of the two galaxy samples. The orange dotted curve represents the GSMF obtained by \citet{bhatawdekar19} for assuming Sersic light profiles in the completeness analysis rather than point sources. The blue shaded areas are the same as in the right panel of Fig.~\ref{fig:GSMF-data-fit}, i.e. the 1$\sigma$ and 2$\sigma$ areas for the reference model.}
    \label{fig:GSMFs-literature}
\end{figure}

\begin{table}
	\caption{Best-fit GSMF Schechter parameters of our reference model and previously published studies.}
	\label{tab:literature_schechter-parameters}
\begin{tabular}{lccc}
\hline
Reference                                                &   $\log M_0$              &   $\alpha$                    &   $\log \phi_0$\\
                                                         &   (M$_{\sun}$)            &                               &   (Mpc$^{-3}$)\\
\hline
This Work\tablenotemark{a}                               &   $10.21_{-0.26}^{+0.43}$ &   $-1.96_{-0.08}^{+0.09}$     &   $-4.48_{-0.31}^{+0.62}$\\
This Work\tablenotemark{b}                               &   $10.22_{-0.27}^{+0.45}$ &   $-1.96_{-0.08}^{+0.09}$     &   $-4.49_{-0.32}^{+0.64}$\\
\citet{song16}\tablenotemark{c}                          &   $10.72_{-0.30}^{+0.29}$ &   $-1.91_{-0.09}^{+0.09}$     &   $-4.86_{-0.24}^{+0.53}$\\
\citet{bhatawdekar19}\tablenotemark{c,d}                 &   $10.29_{-0.67}^{+0.65}$ &   $-1.89_{-0.10}^{+0.09}$     &   $-4.27_{-0.26}^{+0.65}$\\
\citet{bhatawdekar19}\tablenotemark{c,e}                 &   $10.35_{-0.50}^{+0.50}$ &   $-1.98_{-0.07}^{+0.07}$     &   $-4.22_{-0.25}^{+0.64}$\\
\citet{kikuchihara20}                                    &   $9.58_{-0.15}^{+0.23}$  &   $-1.85_{-0.07}^{+0.07}$     &   $-3.74_{-0.22}^{+0.30}$\\
\hline
\end{tabular}
\tablenotetext{a}{Using a Schechter function Eq.~\eqref{eq:schechter}}
\tablenotetext{b}{Using a modified Schechter function Eq.~\eqref{eq:modified_schechter}}
\tablenotetext{c}{$z\sim6$ sample}
\tablenotetext{d}{Point source results}
\tablenotetext{e}{Disk galaxy results}
\end{table}

\begin{figure}
    \centering
    \includegraphics[width=0.5\textwidth, keepaspectratio=True]{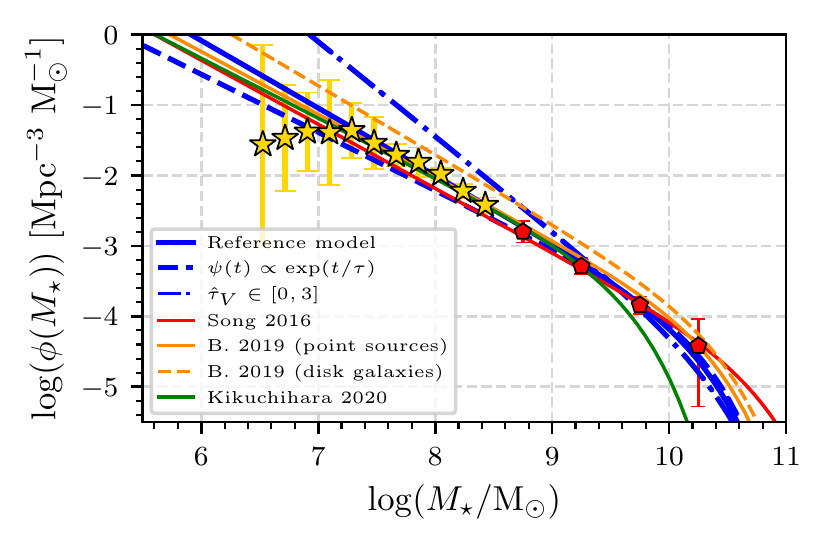}
    \includegraphics{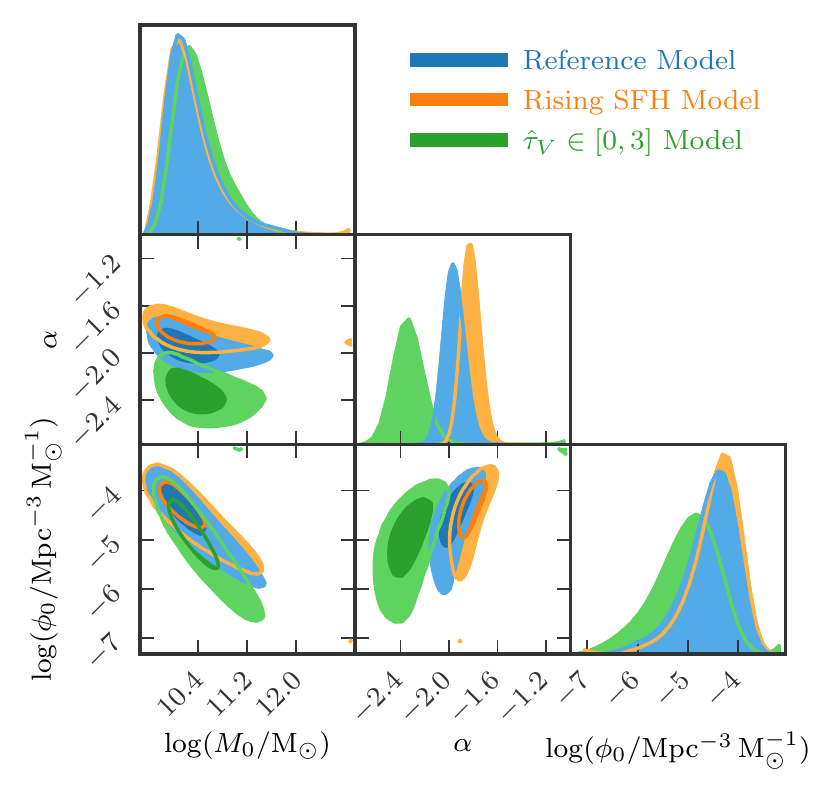}
    \caption{\textit{Upper Panel}: The same plot as Fig.~\ref{fig:GSMFs-literature} with in addition the Schechter-fits of the GSMF points derived for an exponentially rising SFH (\textit{dashed blue curve}) and for 'dusty' galaxies (\textit{dash-dotted blue curve}). The two degeneracies, SFH and dust attenuation, have opposite impacts on the best-fit GSMF. These two effect appear to cancel each other out and result in similar GSMFs as our reference model if both the SFH and allowing a wide range of possible dust attenuation optical depths are included in the modelling simultaneously \citep[e.g.,][]{song16,bhatawdekar19}. \textit{Lower panel:} Posterior distributions of the three Schechter parameters for the reference model, the exponentially rising SFH model and the 'dusty' $\hat{\tau}_V\in[0,3]$ model. These plots illustrate the opposing effects of changing the assumed SFH and allowing the dust attenuation to roam over a wide range have on the low-mass end of the GSMF.}
    \label{fig:GSMFs-literature_models}
\end{figure}

Previous studies of the GSMFs at $z\sim6-7$ include observations of galaxies in both blank fields \citep{gonzales11,duncan14,grazian15,song16} and those lying behind HFF clusters \citep{bhatawdekar19,kikuchihara20}.
The \citet{song16} $z\sim6$ GSMFs are based on a large sample of galaxies and UV luminosity functions derived by \citet{finkelstein15} in the CANDLES/GOODS, the ERS and the HUDF fields and thus represent the most complete study of high-redshift GSMFs in blank fields thus far. Previous studies \citep{gonzales11,duncan14,grazian15} studied sub-samples of the CANDLES/GOODS fields which are all included in the sample studied in \citet{song16}. We will therefore rely mostly on the results of \citeauthor{song16} when comparing blank field results to this work.  \citeauthor{song16} used 12 bands of HST and \textit{Spitzer}/IRAC photometry and include full modelling of the SFH in their SED-fitting as well as allowing a significant range in the dust attenuation, $A_V\in[0,3.2]$ (about the same range as our `dusty' model).

The two high-redshift GSMF studies observed through SL clusters \citep{bhatawdekar19,kikuchihara20} are both based on the 7 HST bands, the \textit{Ks} band and the two deep \textit{Spitzer}/IRAC bands available for the HFF clusters. \citet{bhatawdekar19} derive their $z\sim6$ UV luminosity functions and GSMFs from a sample of galaxies detected in the cluster and parallel field of MACS0416. Their analysis contains two GSMFs for $z\sim6$ galaxies that were derived in two different ways: One assuming that the galaxies all have Sersic profiles light profiles and the other in which they assumed all detections are point sources. These assumptions resulted in two different completeness limits and corrections. They also include full modelling of the SFH and dust attenuation ($A_V\in[0,2]$) in their SED-fitting and their method of treating SL magnification is similar to ours, i.e., they take the median magnification of all SL models available for MACS0416. The GSMFs derived by \citet{kikuchihara20} on the other hand are based on a $z\sim6-7$ galaxy sample detected in all six HFF clusters by \citet{kawamata18} and UV luminosity functions derived by \citet{ishigaki18}. They use the GLAFIC SL models to derive their magnification factors and \texttt{BEAGLE} with a constant SFH and $\hat{\tau}_V\in[0, 2]$ for their SED-fitting. Finally, they also use the \citeauthor{song16} estimates of the co-moving density as a function of stellar mass to constrain the high-mass end of their derived GSMF.

We show our best-fit reference model GSMFs alongside the best-fit results from the literature in Fig.~\ref{fig:GSMFs-literature} and the corresponding best-fit Schechter parameters and their $1\sigma$ uncertainties in Table~\ref{tab:literature_schechter-parameters}. Overall, our results are in good agreement with results in the literature within the uncertainties. Assuming a disk galaxy light distribution to estimate the completeness, \citet{bhatawdekar19} found results that deviate somewhat from the consensus of the results from the literature; their results when they assumed the sources had point-like light distributions in deriving the completeness agree well with ours. This is unsurprising since our $z\sim6-7$ sample is mostly comprised of unresolved sources. At the low-mass end, we find a slope of $\alpha\simeq-1.96_{-0.08}^{+0.09}$ which is slightly steeper than in the literature but agrees with \citet{song16} and \citet{bhatawdekar19} within 1$\sigma$. \citet{kikuchihara20} on the other hand find a shallower low-mass end slope of $\alpha\simeq-1.85_{-0.07}^{+0.07}$ which differs from our estimated slope at about the 2$\sigma$ level. We also note a slight discrepancy between \citeauthor{song16}, \citeauthor{kikuchihara20} and our own results at the high-mass end even though these three studies use the same high-mass GSMF results. This is due to the role that the exact properties of the low-mass end GSMF play in constraining the overall shape of the GSMF. The high-mass end exponential cutoff mass, $M_0$, and the low-mass end slope, $\alpha$, of the GSMF are not independent parameters as illustrated by the shape of their posterior distributions in Fig.~\ref{fig:corner_reference-model}. This work is the only study of $z\sim6-7$ GSMFs to date that observes a possible turnover of the GSMF at very low masses at greater than 1$\sigma$-level. Note that \citet{kikuchihara20} and our work are currently the only studies that probe stellar masses as low as $M_{\star}\gtrsim10^{6}\,\text{M}_{\sun}$ at $z\sim6-7$.

Our reference model GSMF results and the literature broadly agree within the uncertainties despite the fact that the different results are based on different sets of assumptions (Fig.~\ref{fig:GSMFs-literature}). This does not contradict our conclusions of the importance of parameter choice and assumptions in SED-fitting (section~\ref{sec:degeneracies}): The degeneracy between $M_{\star}$ and $\hat{\tau}_V$ results in a steeper low-mass end slope of the GSMF than the reference model and adopting an exponentially rising SFH in a shallower low-mass end slope (see Fig.~\ref{fig:GSMFs}). When modelling \textit{both} SFH and dust attenuation in the same SED-fit, as is done in \citet{song16} and \citet{bhatawdekar19}, these two degeneracies are both present in the fits. However, each of the SFHs and dust attenuations assumed in both studies tend to push the slope in opposite directions and thus effectively and coincidentally cancel each other out, yielding average GSMFs similar to our reference model (Fig.~\ref{fig:GSMFs-literature_models}). While \citet{kikuchihara20} also use a constant SFH in their SED-fitting, the degeneracy between dust attenuation and metallicity, which is a free parameter in their study, has a similar effect on their resulting GSMF. While the reference model agrees with the results from the literature, different SFHs and constraints on dust attenuation can lead to significantly different GSMFs. Because of this, it is crucial to explore the impact of different SED-fitting assumptions on the high-redshift GSMF in understanding the robustness of any GSMF.

\subsection{Theoretical implications} \label{sec:theory}

Robust determinations of the form the UV luminosity and stellar mass functions of the first generations of galaxies provide deep insight into and strong constraints on the physical processes which drive galaxy evolution at all redshifts. Given the importance of these determinations,  theorists and simulators have expended great effort in predicting UV luminosity and sometimes mass functions of galaxies at $z\ga6$ \citep[see][and references therein for a review]{bromm11}. There is a general agreement in theory and simulations that both the ionization of the gas within and falling onto galaxy halos by the meta-galactic flux responsible for reionization and feedback from massive stars and perhaps AGN play key roles in shaping the low mass and luminosity ends of the stellar mass and UV luminosity functions \citep{gnedin16, liu16, finlator18, ocvirk20}. Modeling correctly how the UV continuum luminosity and color of any galaxy relate to its stellar (and DM halo) mass is obviously crucial to making the link between the UV luminosity and stellar mass functions at any redshift but particularly at high redshift when galaxies are on average quite young and are only selected by the intensity of their UV continuum or line emission. Several attempts at making a direct comparison between stellar mass and UV luminosity in observations and simulations have been published \citep{dayal14, duncan14, o'shea15, yue16,liu16,hutter20}.

The various simulations results that have been published have somewhat different UV luminosity and stellar mass functions. All of them lead to the decrease in the number density of low luminosity and low mass halos (or stellar masses). The virial temperature, $T_{\rm vir}$, of a DM halo at $z\sim6-7$ of mass 10$^8$\,\Msun\ is about 10$^4$\,K. Thus if all of the hydrogen is ionized by the meta-galactic flux in halos of this mass, halos below this mass will not be able to retain their baryons. This is the ultimate limit for halos to retain their accreted gas mass and thus, it is likely that halos with masses less than this limit will have strongly suppressed star formation. Including the effects of feedback in the form of heating of the interstellar and circum-galactic gas and driving outflows increases the halo masses at which baryons are ejected. Simulations that include the effects of feedback from massive stars have declines in co-moving number density in their simulated luminosity functions (and in some studies stellar mass functions) at higher halo masses than the limit
calculated through the ionization heating limit when $T_{\rm vir}=10^4$\,K. For example, \citet{ocvirk20} find that star formation is gradually suppressed below a DM halo mass of $2\times10^9$\,\Msun\ which corresponds to a turnover in the luminosity function at $M_{UV}=-11$ in their models. Other simulations predict a similar turnover or flattening in the luminosity function \citep[e.g.,][]{yue16} or at a magnitude or two higher UV luminosities \citep[e.g.,][]{gnedin16,liu16}. It is difficult to relate these turnover or flattening magnitudes to the stellar mass but in some of the simulations, the flattening occurs for relatively small DM halo masses \citep{liu16} compared to others, e.g, \citet{ocvirk20}.

If galaxies in the early universe are forming inefficiently (low ratios of mass accretion rates to SFRs), then we expect young galaxies to have relatively low SFRs and thus relatively high ratios of DM mass to stellar masses \citep[$\sim0.1$\,\% or less at $z\sim6$; see, e.g.,][]{behroozi13}. We can make a crude estimate of the mass where energy injection rate of massive stars through stellar winds and supernova explosions is sufficient to keep galaxies from forming stars efficiently. A strong criterion for doing this would be the SFR at which the thermalized energy from stellar winds and supernovae becomes about the binding energy of the galaxy halo.  Using Eqn.~2 from \citet{bromm11}, we can estimate the binding energy of a halo as, 

\begin{equation} \label{eq:twobromm11}
    E_{\rm binding}\sim10^{57} \left(\frac{M_{\rm vir}}{10^{10}\,\text{M}_{\odot}}\right) \left(\frac{\Delta_c}{18\pi^2}\right) \left(\frac{1+z}{10}\right)\,\rm erg
\end{equation}

\noindent 
where $M_{\rm vir}$ is the virial mass, $\Delta_c$ is the overdensity of a virialized halo, and $z$ is the redshift. Setting $z=6$ and $\Delta_c=18\pi^2$, results in $E_{\rm binding}\sim
7\times10^{56}(M_{\rm vir}/10^{10}$\,M$_{\odot}$)\,erg. The mechanical energy from young stars through their stellar winds and supernovae (SNe) can be estimated as,

\begin{equation} \label{eq:Ewind}
    E_{\rm wind}\sim3\times10^{56} \left(\frac{\psi}{\text{M}_{\odot}\,\text{yr}^{-1}}\right) \epsilon_{\rm therm} \left(\frac{t_{SF}}{10^7\,\text{yr}}\right)\,\rm erg
\end{equation}

\noindent
where $\psi$ is the star formation rate in \Msunyr, $\epsilon_{\rm therm}$ is the thermalization efficiency defined as the ratio of total outflow energy and the mechanical energy output by young stars averaged over the duration of the (burst of) star formation, $t_{\rm SF}$. To scale the relation between the SFR and the mechanical energy output from young stars, we used 10$^{42}$\,erg\,s$^{-1}$ which is the approximate instantaneous mechanical energy output at $\sim$few 10$^7$\,yr and longer for a constant SFR of a solar mass per year \citep{SB99}. We can recast this equation because of our assumption of a constant SFR in our reference model such that $E_{\rm wind}\sim3\times10^{56}(M_{\star}/10^7\,$M$_{\odot})\,\epsilon_{\rm therm}\,(t_{SF}/10^7$\,yr)\,erg. Combining these two equations to estimate the ratio of the wind energy to binding energy of a halo we find,

\begin{equation} \label{eq:ratio}
   \begin{split}
    \frac{E_{\rm wind}}{E_{\rm binding}} & \sim0.4 \left(\frac{M_{\star}}{10^7\,\text{M}_{\odot}}\right) \epsilon_{\rm therm} \left(\frac{M_{vir}}{10^{10}\,\text{M}_{\odot}}\right)^{-5/3} \left(\frac{t_{SF}}{10^7\,\text{yr}}\right) \\
    & =0.4 \left(\frac{M_{\star}}{M_{vir}}\right)_{0.001} \epsilon_{\rm therm} \left(\frac{M_{vir}}{10^{10}\,\text{M}_{\odot}}\right)^{-2/3} \left(\frac{t_{SF}}{10^7\,\text{yr}}\right)
   \end{split}
\end{equation}

\noindent
where ($M_{\star}$/$M_{vir}$)$_{0.001}$ is the ratio of the stellar to virial mass in units of 10$^{-3}$. Since the stellar mass of a galaxy is approximately linearly proportional to its SFR times its age under the assumption of a constant SFH (and from the empirical relation of stellar mass and SFR), the energy of outflows at constant thermalization efficiency will increase more slowly, only linearly, than the binding energy of the halo as a function of halo mass which increases to the power 5/3. While we do not know the stellar mass to halo mass ratio of galaxies with stellar mass of 10$^7$\,\Msun\ at $z\sim6$, it is likely to be of order 0.01 \citep[e.g.,][]{behroozi13}. If the stellar to halo mass ratio declines less rapidly than $M_{\text{vir}}^{2/3}$, then the ratio of the wind energy to the binding energy will also increase more rapidly as the halo mass decreases. In the event of either constant or slowly declining $M_{\star}/M_{\text{vir}}$ with decreasing halo masses, galaxy formation will become increasingly inefficient with decreasing stellar mass below halo masses of about a few 10$^9$\,\Msun~\citep[depending of course on the thermalization efficiency, which has been estimated to be $\sim0.5$;][]{strickland09}. We emphasize that the parameter choices we made for this estimate are somewhat tuned for low halo masses and given the complex physics of outflows, this estimate cannot be simply extrapolated to other galaxy stellar and halo masses. Although crude, the estimate is consistent with the simulations of \citet{ocvirk20} in that galaxies residing in halos of masses $\sim$few 10$^{9}$\,\Msun\ or less will form inefficiently. The results of some simulations and our crude analysis suggest that feedback provided by stellar winds and SNe may provide sufficient energy to keep the formation of galaxies in low mass halos inefficient if the thermalization efficiency is relatively high \citep{strickland09}. Observations with JWST will provide a deeper understanding as to whether or not feedback is in fact the culprit in keeping galaxy formation inefficient at high redshift and, in particular, for low-mass galaxies near this mass-turnover.

\section{Summary and Conclusion} \label{sec:conclusion}

We presented the derivation of high-redshift $z\sim6-7$ galaxy stellar mass functions under various spectral energy distribution fitting assumptions using a sample of gravitationally lensed galaxies lying behind the six HFF clusters. For the sample of 303 $z\sim6-7$ galaxies selected via the drop-out technique from deep HST imaging data, we measured the \textit{Ks} and \textit{Spitzer}/IRAC photometry and computed median gravitational magnification factors from the SL models available for the HFF clusters. We derived stellar masses by fitting 10 broad-band photometry filters with the \texttt{BEAGLE} SED-fitting tool. In order to test the impact of various SED-fitting assumptions on the resulting GSMF, we first constructed a minimum-parameter model which assumed a constant SFH and a fixed metallicity, $Z=0.1\,\text{Z}_{\sun}$, as a reference model. To test how robust the GSMF to changing assumptions in the SFH, metallicity, and range over which the amount of dust attenuation is allowed to roam, we ran SED fitting models with different assumptions for metallicity, the functional forms of the star formation histories, and both the type and range in the amount of dust attenuation allowed. The derived stellar masses were then used to derive $M_{\star}-M_{UV}$-scaling relations which allowed us to convert the rest-frame UV luminosity function of our galaxy sample, derived by \citet{atek18}, to galaxy stellar mass functions. Finally, we determined best-fitting GSMFs and their uncertainty ranges by exploring the parameter space with an MCMC analysis.

The summary and main conclusions of our analysis are:

\begin{itemize}
    \item We extend the high-redshift GSMF to low masses, $M_{\star}>10^{6}$\,M$_{\sun}$. The best fit parameters for the $z\sim6-7$ GSMF are a high-mass exponential cutoff at $\log(M_0/\text{M}_{\sun})\simeq10.22_{-0.27}^{+0.45}$, a relatively steep low-mass end slope, $\alpha\simeq-1.96_{-0.08}^{+0.09}$, and a normalization $\log(\phi_0/\text{Mpc}^{-3})\simeq-4.49_{-0.32}^{+0.64}$ for our reference model (Table~\ref{tab:SED-fit_parameters}). These results are in good agreement with recent results from the literature within the respective error bars.
    
    \item The $z\sim6-7$ GSMF departs from the Schechter form at the very low-mass end with a downward turnover at $\log(M_T/\text{M}_{\sun})\simeq7.10_{-0.56}^{+0.17}$ and a curvature $\beta\simeq1.00_{-0.73}^{+0.87}$ in the reference model. While the uncertainties on the lowest-mass bins also allow for an upward turnover of the GSMF, most models do not have an upturn and this scenario is ruled out at the greater than 1$\sigma$-level by our MCMC analysis for the reference model and most SED-fitting configurations.
    
    \item Due to large systematic differences between the various SL models available for the HFF clusters, the gravitational magnification factors are uncertain but are especially high for galaxies with high median magnification factors. While gravitational lensing is very useful for \textit{detecting} faint high-redshift objects, one needs to fully account for lensing uncertainties. Therefore, unless lensing models are better constrained, especially the caustics close to the center of the cluster potential and the BCGs, the utility of using strongly lensed galaxies to probe the low mass end of the GSMF is limited.
    
    \item Obtaining accurate rest-frame optical IRAC photometry, vital for constraining stellar masses of individual galaxies in this type of study, proves very challenging when observing very crowded fields such as the HFF. Using a subset of galaxies with reliable IRAC photometry, we empirically estimate a correction factor for stellar masses based on using only HST photometry which probes only the rest-frame UV.
    
    \item While not strongly affected by metallicity, the determination of high-redshift GSMFs depends significantly on assumptions about the SFH and dust attenuation. For example, a significantly shallower low-mass end slope, $\alpha\simeq-1.82_{-0.07}^{+0.08}$, is the best fit for an exponentially rising SFH, $\psi(t)\propto\exp{(t/\tau)}$. When allowing for a wider range and higher limit on the amount of dust attenuation, $\hat{\tau}_V\in[0,3]$, we find a significantly steeper low-mass end slope, $\alpha\simeq-2.34_{-0.10}^{+0.11}$.
    
    \item Our results show that variations in the slope of the $M_{\star}-M_{UV}$-relation and in $M_{\star}$ of the order of $\gtrsim0.1$\,dex can significantly impact the resulting GSMFs. Fluctuations of this magnitude arise both from SED-fitting assumptions and the lack of rest-frame optical detections or sensitive upper-limits in the IRAC bands for all galaxies.
    
    \item The Schechter parameters of the GSMF are not generally independent. In particular, there is a degeneracy between the high-mass exponential cutoff, $M_0$, and the low-mass end slope, $\alpha$. In order to break these degeneracies and to robustly constrain the overall shape and parameters of the GSMF, detections of galaxies at both high and low stellar masses are required.
\end{itemize}

The derivation and study of high-redshift GSMFs remain susceptible to large uncertainties, parameter degeneracies, and on the specific assumptions made when fitting their SEDs. Our approach of converting rest-frame UV luminosity functions to high-redshift GSMFs does not take into account the non-trivial problem of stellar mass incompleteness, which is not necessarily the same as UV luminosity incompleteness. Future studies will need to address this issue in order to derive robust $z\sim6-7$ GSMFs, in particular considering how sensitive the GSMF is to fluctuations in the $M_{\star}-M_{UV}$-relation used to convert the UV luminosity function to the stellar mass function. Considerable advances in SL modelling techniques, deeper images at more wavelengths which probe the rest-frame optical in distant galaxies, and finding and determining redshifts of multiply lensed galaxies in the cores of the HFF clusters to improve the lensing models are also all required to better constrain the very low-mass end of the GSMF. Deep observations of high-redshift galaxies with ALMA will yield new insights and constraints on the dust properties of these objects and thus help to break one of the major degeneracies in SED fitting. The greatest advancements in the study of very high-redshift galaxies will however come with the launch of the JWST. JWST will provide the first rest-frame optical and NIR spectra of galaxies at $z\gtrsim6$ and thus the first robust constraints on stellar masses of low mass galaxies. The very faint and low-mass $z\sim6-7$ galaxies studied in this work are prime targets for follow-up with the JWST.

\section*{Acknowledgements}

The authors warmly thank the anonymous referee for her or his very useful and constructive comments which greatly helped to improve the paper. L.F. in particular thanks Amanda Pagul, Iary Davidzon, John Weaver, Marko Shuntov and the COSMOS2020 team for some very interesting and insightful discussions about \textit{Spitzer}/IRAC photometry and SED-fitting. L.F. also thanks Lucas Pinol and Guilhem Lavaux for equally insightful and crucial exchanges about likelihoods and MCMC-modeling. This work is supported by CNES. J.C. and S.C. acknowledge financial support from the European Research Council (ERC) via an Advanced Grant under grant agreement no. 321323--NEOGAL. This work is based on observations obtained with the NASA/ESA \textit{Hubble Space Telescope} (HST), retrieved from the \texttt{Mikulski Archive for Space Telescopes} (\texttt{MAST}) at the \textit{Space Telescope Science Institute} (STScI). STScI is operated by the Association of Universities for Research in Astronomy, Inc. under NASA contract NAS 5-26555. This work utilizes gravitational lensing models produced by PIs, Natarajan \& Kneib (CATS), Sharon, Williams, Keeton, Bernstein and Diego, and the GLAFIC group. This lens modeling was partially funded by the HFF program conducted by STScI. The lens models were obtained from the \texttt{MAST} archive. This work is based in part on data and catalog products from HFF-DeepSpace, funded by the National Science Foundation and STScI. This work has made use of the \texttt{CANDIDE} Cluster at the \textit{Institut d'Astrophysique de Paris}, made possible by grants from the PNCG and the region of Île de France through the program DIM-ACAV+. This research made use of \texttt{Astropy},\footnote{\url{http://www.astropy.org}} a community-developed core Python package for Astronomy \citep{astropy13,astropy18} as well as the packages \texttt{NumPy} \citep{vanderwalt11} and \texttt{SciPy} \citep{virtanen20}. \texttt{Matplotlib} \citep{hunter07} and the \texttt{pygtc} package \citep{bocquet16} were used to create the figures in this work.

\section*{Data availability}

The data and lensing models underlying this article are publicly available in the \texttt{MAST} archive: \url{https://archive.stsci.edu/pub/hlsp/frontier} and in the NASA/IPAC \texttt{IRSA} archive: \url{https://irsa.ipac.caltech.edu/data/SPITZER/Frontier}.




\bibliographystyle{mnras}
\bibliography{references} 



\appendix

\section{Stellar age constraints in SED-fitting} \label{app:age-impact}

\begin{figure}
    \centering
    \includegraphics{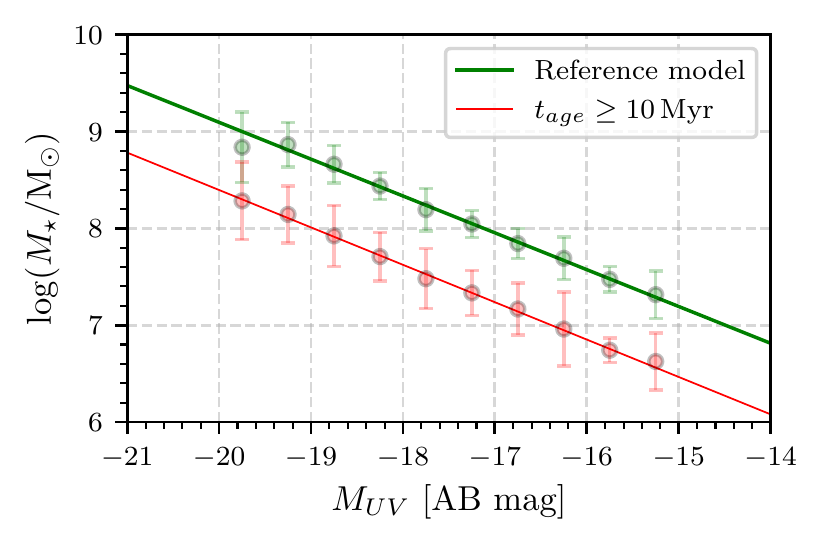}
    \caption{Best-fit $M_{\star}-M_{UV}$-relations for a model that allows stellar ages down to 10\,Myr (red) and the reference model (green). The two relations have the same slope, $a\sim-0.38$, but allowing stellar ages down to 10\,Myr yields considerably lower stellar masses (by $\sim0.6$\,dex).}
    \label{fig:10Myr_M-Muv_relation}
\end{figure}

\begin{figure}
    \centering
    \includegraphics{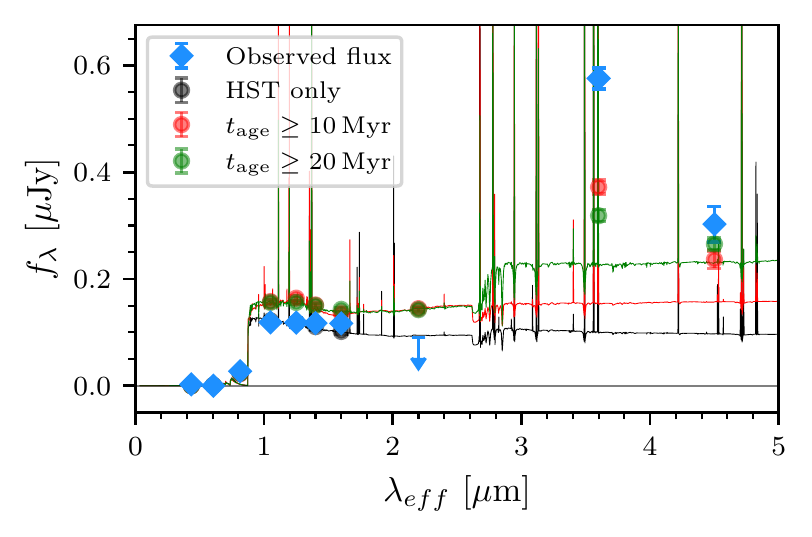}
    \caption{Median-stacked best-fit SEDs of sources with reliable IRAC photometry at the median redshift of the golden and silver sources $z_{\text{med}}=6.28$. The blue squares represent the median observed fluxes in each band. The colored circles represent the maximum \textit{a posteriori} fluxes of the stacked best-fit SEDs in each band. The green SED is for the reference model, the red SED allows for ages down to 10\,Myr and the black SED is fit only to the seven HST bands and illustrates how the stellar mass in underestimated when only fitting rest-frame UV photometry. Note that since the majority sources are beyond the detection limit in the \textit{Ks} band, we plot the upper limiting flux for the \textit{Ks} band here.}
    \label{fig:median-stacked_SED}
\end{figure}

\begin{figure}
    \centering
    \includegraphics{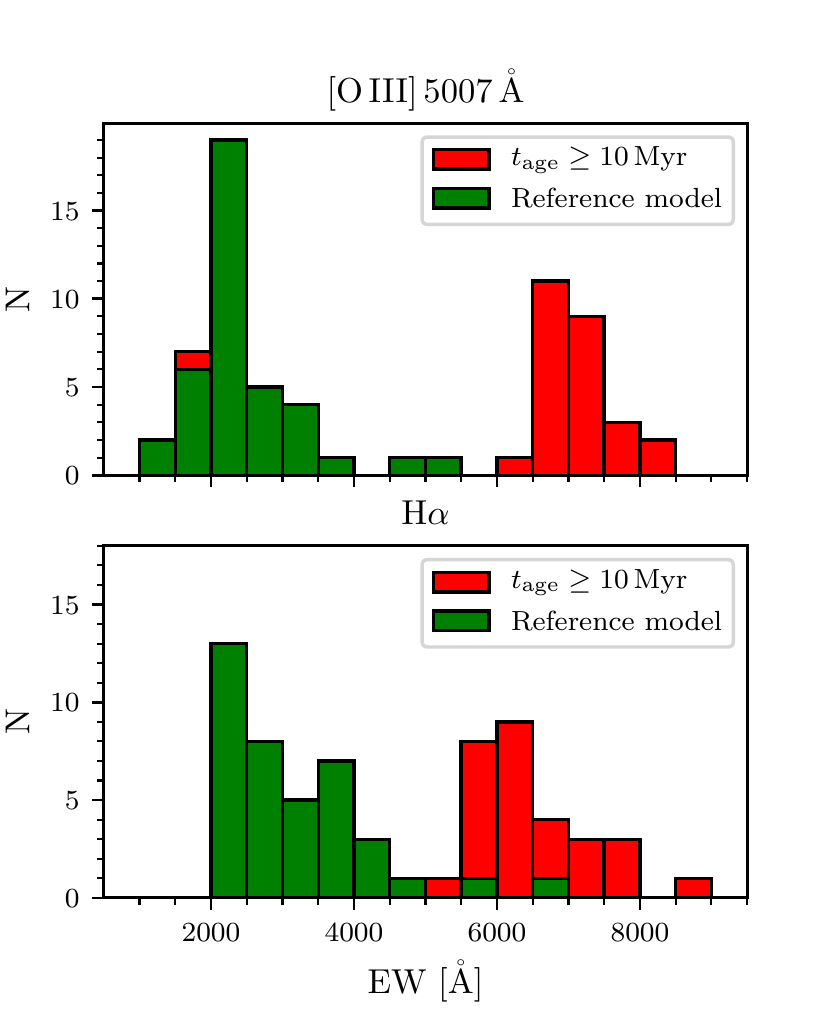}
    \caption{Best-fit (maximum-a-posteriori) nebular EW distribution of [\ion{O}{iii}]\,5007\,\AA~(\textit{upper panel}) and H$\alpha$ (\textit{lower panel}) from our SED-fit with \bea\ for all golden and silver sources. EWs for the reference \bea~configuration are shown in green and the EWs for a model that allows stellar ages down to 10\,Myr are shown in red.}
    \label{fig:line-EWs}
\end{figure}

The estimate of stellar mass from SED-fitting is tightly correlated with stellar age. As described in section~\ref{sec:SED-fitting}, we place dynamical constraints on the range of allowed maximum stellar ages with a lower boundary of 20\,Myr. To illustrate the impact of the allowed range of stellar age on the SED-fit, we ran a model which allows stellar ages as low as 10\,Myr with the otherwise same parameter configuration as the reference model (including a flat initial distribution of the age priors). The resulting best-fit SEDs (i.e., those with the maximum in the \textit{posterior} maximum likelihood probability distribution) tend towards extremely young, $\sim10$\,Myr, galaxies with masses lower by $\sim0.6$\,dex on average compared to the average values of the reference model (e.g., $\log(M_{\star}/\text{M}_{\odot})_{(M_{UV}=-19.5)}\simeq8.20\pm0.07$). The slope of the $M_{\star}-M_{UV}$-relation remains the same in the two models (cf. Fig.~\ref{fig:10Myr_M-Muv_relation}). The consequences for the GSMF are therefore similar to adopting an exponentially rising SFH (cf. section \ref{sec:SFH-impact} and Fig.~\ref{fig:GSMF-rising_SFH}) though more pronounced: A shallower low-mass end slope $\alpha$ and a lower turnover mass $M_T$. 

These much lower stellar masses are an effect of nebular emission in the SED-fitting analysis: As can be seen in Fig.~\ref{fig:median-stacked_SED}, in the case where we allow stellar ages down to 10\,Myr (red) the best-fitting SEDs have a much lower rest-frame optical continuum than the reference model (green) but nevertheless \textit{higher} total flux in the IRAC1 band. Indeed, the best-fit SEDs of golden and silver sources (i.e., sources with trusted IRAC photometry) tend to have EWs of $\sim7000$\,\AA~for both the [\ion{O}{iii}]\,5007\,\AA~line, which is typically redshifted to the IRAC1 band in our $z\sim6-7$ sample, and the H$\alpha$ line, typically redshifted to the IRAC2 band (cf. Fig. \ref{fig:line-EWs}). The Bayesian fit as performed by \texttt{BEAGLE} therefore seems to favor rest-frame optical photometry dominated by nebular emission rather than stellar continuum emission.

Though these extremely high nebular EWs and subsequent young stellar ages represent maximum likelihood ('best-fit') solutions in the Bayesian fitting procedure, a closer look at the posterior distribution of the SED-fit parameters reveals the stellar age distribution to be bi-modal for most of the sources with IRAC photometry: A primary (maximum likelihood) peak close to $\sim10$\,Myr and a secondary peak around $100$\,Myr. While the likelihood of the fit as estimated by \texttt{BEAGLE} favors the younger solution, observations favor the older solution since nebular EWs of $\sim7000$\,\AA~have not yet been observed. Spectroscopic observations of low-redshift analogs of high-redshift galaxies merely found nebular emission line EWs $<2000$\,\AA~\citep{atek11,atek14b,reddy18b}, which roughly corresponds to the order of magnitude of best-fit EWs in the reference model (cf. Fig.~\ref{fig:line-EWs}). While it is indeed possible to find single very young galaxies of $\lesssim10$\,Myr at $z\sim6-7$ \citep[e.g.][]{ono10,mclure11}, they are unlikely to represent our whole sample. These objects are furthermore modeled as single starbursts \citep{mclure11} which is not compatible with the continuous (constant and exponential) SFHs adopted in our SED-fitting technique. A small sample of bright and massive $z\sim6-7$ galaxies has however recently been modeled with a continuous SFH and an ongoing starburst yielding very large optical EWs \citep[up to $6240_{-3450}^{+1540}$\,\AA\ for {[\ion{O}{iii}]}+H$\mathrm{\beta}$ for one source;][]{endsley21}. The median UV continuum slope over our $z\sim6-7$ sample $\beta_{UV}\sim-2.1$ further supports stellar ages $\gtrsim100$\,Myr given the metallicities and SFHs that we probe rather than younger objects which would be expected to have bluer UV slopes \citep{dunlop13}.

For these reasons, we favor the older $\sim100$\,Myr solutions over the young maximum likelihood solutions. We therefore choose to place a lower boundary of 20\,Myr on the maximum stellar age in our SED-fitting procedure in order to exclude these extremely young solutions from the overall likelihood estimates. Future observations of the rest-frame optical emission lines with JWST/NIRSpec will be required to measure the the EWs of the nebular emission lines and disentangle the nebular emission from the stellar continuum in order to robustly constrain stellar age and mass at $z\sim6-7$ and probe possible very young starbursts.

\section{Biases in correcting for missing IRAC photometry} \label{app:irac-correction_bias}

\begin{figure}
    \centering
    \includegraphics[width=0.5\textwidth, keepaspectratio=True]{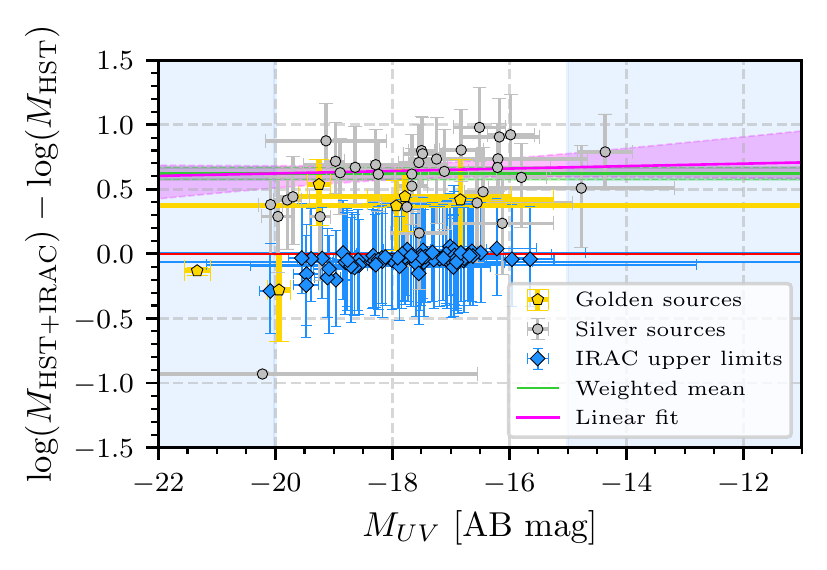}
    \caption{Same plot as in Fig.~\ref{fig:IRAC-photometry_correction} with in addition a linear fit to the golden and silver sources (excluding the three bright outliers described in section~\ref{sec:IRAC-photometry_correction}) shown as a magenta line and its 1$\sigma$-range as the magenta shaded area. The fit is clearly consistent with the constant correction factor $\langle\delta\rangle$ derived in section~\ref{sec:IRAC-photometry_correction} (green line) over the effective $M_{UV}$-range probed in this analysis ($-20\leq M_{UV}\leq-15$, delimited by the blue shaded areas as in Fig.~\ref{fig:m-muv-relation_data}).}
    \label{fig:IRAC-photometry_correction_fit}
\end{figure}

\begin{figure}
    \centering
    \includegraphics[width=0.5\textwidth, keepaspectratio=True]{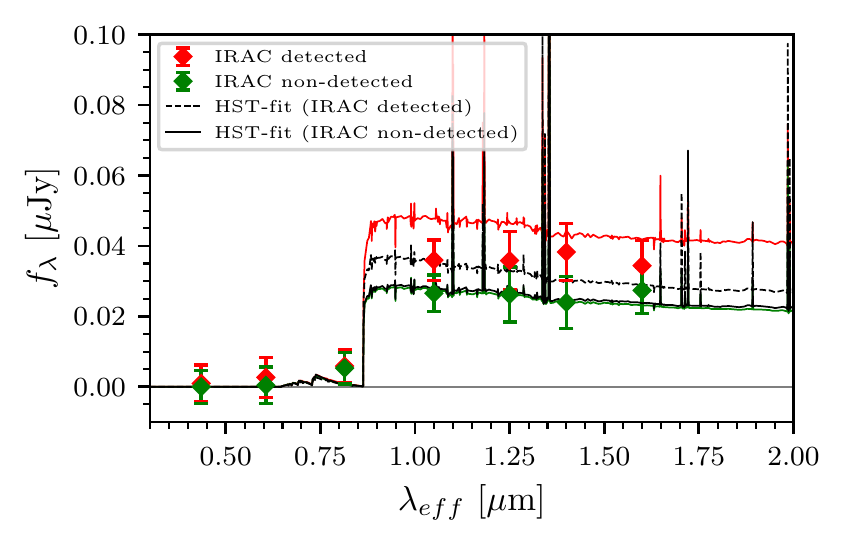}
    \caption{Median-stacked best-fit SEDs of sources detected in at least one IRAC band (golden and silver sources; red line) and of sources that are not detected by the 3$\sigma$-limits in both IRAC1 and IRAC2 (green line) at the median redshift $z_{\text{med}}=6.08$. The colored pentagons represent the corresponding median flux in each band. The dashed and solid black lines show the median-stacked best-fit SEDs when only fitting HST-photometry for both galaxy samples respectively. The SEDs were trimmed down to only show the rest-frame UV emission probed by the HST-bands.}
    \label{fig:IRAC-photometry_SEDs}
\end{figure}

As described in section~\ref{sec:IRAC-photometry_correction}, we correct the stellar masses for sources without reliable \textit{Spitzer}/IRAC photometry either because they are contaminated with light from other galaxies or because they only have upper limits, using an empirical correction factor. We derived a constant correction factor $\mathbf{\langle\delta\rangle}\simeq0.62\pm0.05$\,dex by comparing the stellar masses obtained from fitting SEDs to both HST+IRAC photometry and to HST photometry only. In order to quantitatively asses the applicability of the correction factor over the range of rest-frame UV luminosities probed in this study, we fit a linear relation to the mass offsets of the golden and silver sources as a function of $M_{UV}$, excluding the three very luminous outliers mentioned in section \ref{sec:IRAC-photometry_correction}. The fit is performed with $10^6$ MCMC steps and the result can be seen in Fig.~\ref{fig:IRAC-photometry_correction_fit}. We find a best-fitting slope of $0.01_{-0.04}^{+0.01}$, which is consistent with a constant offset over the whole $M_{UV}$-range probed in this analysis. Our constant correction factor is therefore applicable to our whole sample of galaxies without reliable IRAC photometry.

There is a possibility for this correction to be slightly biased towards higher stellar masses for any given rest-frame UV luminosity (obtained through HST photometry) because a rest-frame optically bright, probably more massive, galaxy is more likely to be detected in at least one of the IRAC bands. We show the median stacked best-fit SEDs of all IRAC detected galaxies and all IRAC non-detected galaxies with clean photometry (upper limits) in Fig.~\ref{fig:IRAC-photometry_SEDs} along with the median stacked best-fit SEDs derived including only HST photometry. The figure shows that the galaxies detected in at least one of the IRAC bands (red) are on average also brighter in their rest-frame UV magnitudes than galaxies not detected in IRAC (green). The best-fitting SEDs using only the HST photometry of the IRAC detected galaxies (black dashed curve) therefore also have higher median flux densities than the HST-only fits to the IRAC non-detected galaxies. This shows that while the HST photometry is not suited to constrain the stellar mass alone, it is sensitive enough to the stellar mass to take a possible difference in average stellar mass of the two samples into account and thus at least mitigate a possible bias towards higher mass galaxies in the constant correction factor $\langle\delta\rangle$. Fig.~\ref{fig:IRAC-photometry_SEDs} also shows that the stacked best-fit SEDs to HST+IRAC upper limits and the stacked best-fit SEDs to only HST photometry are identical. This demonstrates that the upper limits on the flux in both IRAC bands are not sufficient to constrain stellar mass any better than HST photometry alone (see also Fig.~\ref{fig:IRAC-photometry_correction} and Fig.~\ref{fig:IRAC-photometry_correction_fit}).

We do not detect any flux in the stacked IRAC frames of the sources with only upper limits in the IRAC bands. This is not surprising however since we are observing exceptionally crowded fields where much brighter foreground sources lying near our sources and the ICL both contaminate the IRAC fluxes of high-redshift galaxies. We note that in particular, sources flagged as 'silver' are not completely isolated in the IRAC frames, we merely estimate that the aperture in which we measure the IRAC photometry is not significantly contaminated by either sources of contaminating light (cf. section~\ref{sec:ir-Data}). However, in the stack, this extra foreground light leads to unexpectedly high noise levels such that we do not gain a factor $\sqrt{N}$ in signal-to-noise through stacking. Furthermore, stacking sources at different locations over several strong lensing clusters neglects the fact that the magnification and its uncertainties depend on the source position. The stacked flux cannot be simply interpreted as saying much about the physical properties of a given population of galaxies. In the literature, a meticulous stacking analysis conducted in blank fields found low signal-to-noise galaxies to be indeed biased towards higher stellar masses \citep{song16}. In a discussion on systematics, \citet{behroozi19} however argue that these results could also be due to SED-fitting degeneracies (e.g., degeneracy between stellar mass and stellar age). We therefore conclude that our average correction factor $\langle\delta\rangle$ is applicable over the range of UV luminosities probed in our sample of galaxies without IRAC photometry, even though we cannot \textit{completely} rule out a bias towards slightly higher stellar masses with the available data.

\section{Further GSMF material} \label{app:GSMF-points}

\begin{table*}
	\caption{Stellar mass bins of our final $z\sim6-7$ GSMFs computed from UV luminosity bins by \citet{atek18} using the $M_{\star}-M_{UV}$-relations for all \texttt{BEAGLE} configurations in table \ref{tab:M-Muv-Relations}.}
	\label{tab:GSMF_points}
\begin{tabular}{c|cc|ccc|cc}
\hline
\textit{Reference model}   &   \multicolumn{2}{c}{\textit{Metallicity tests}}  &   \multicolumn{3}{c}{\textit{SFH tests}}  &   \multicolumn{2}{c}{\textit{Dust tests}}\\
\hline
$\log M_{\star}$  &   $\log M_{\star}$      &   $\log M_{\star}$    &   $\log M_{\star}$   &   $\log M_{\star}$    &   $\log M_{\star}$    &   $\log M_{\star}$             &   $\log M_{\star}$ \\
                  &    (0.01\,Z$_{\sun}$)   &   (0.5\,Z$_{\sun}$)   &   (delayed)          &   (rising)            &    (declining)        &    ($\hat{\tau}_V\in[0,3]$)    &   (SMC law)\\
M$_{\sun}$        &   M$_{\sun}$            &   M$_{\sun}$          &   M$_{\sun}$         &   M$_{\sun}$          &   M$_{\sun}$          &   M$_{\sun}$                   &    M$_{\sun}$\\
\hline
$8.43$  &   $8.36$  &   $8.55$  &   $8.46$  &   $8.16$  &   $8.68$  &   $8.67$  &   $8.49$\\
$8.24$  &   $8.17$  &   $8.36$  &   $8.26$  &   $7.97$  &   $8.50$  &   $8.58$  &   $8.36$\\
$8.05$  &   $7.98$  &   $8.17$  &   $8.07$  &   $7.79$  &   $8.32$  &   $8.48$  &   $8.22$\\
$7.86$  &   $7.79$  &   $7.99$  &   $7.87$  &   $7.61$  &   $8.15$  &   $8.39$  &   $8.09$\\
$7.67$  &   $7.60$  &   $7.80$  &   $7.68$  &   $7.43$  &   $7.97$  &   $8.30$  &   $7.96$\\
$7.48$  &   $7.42$  &   $7.61$  &   $7.48$  &   $7.25$  &   $7.80$  &   $8.20$  &   $7.83$\\
$7.29$  &   $7.23$  &   $7.42$  &   $7.29$  &   $7.06$  &   $7.62$  &   $8.11$  &   $7.69$\\
$7.10$  &   $7.04$  &   $7.23$  &   $7.10$  &   $6.88$  &   $7.44$  &   $8.01$  &   $7.56$\\
$6.91$  &   $6.85$  &   $7.05$  &   $6.90$  &   $6.70$  &   $7.27$  &   $7.92$  &   $7.43$\\
$6.72$  &   $6.66$  &   $6.86$  &   $6.71$  &   $6.52$  &   $7.09$  &   $7.82$  &   $7.29$\\
$6.53$  &   $6.47$  &   $6.67$  &   $6.51$  &   $6.34$  &   $6.91$  &   $7.73$  &   $7.16$\\
\hline
\end{tabular}
\end{table*}

\begin{table*}
	\caption{Final $z\sim6-7$ GSMF bins $\phi(M_{\star})$ and their uncertainties for each \texttt{BEAGLE} configuration. The first column shows the mass bins of the reference model.}
	\label{tab:GSMF_errors}
\begin{tabular}{cc|c|cc|ccc|cc}
\hline
    &   &   \textit{Reference model}   &   \multicolumn{2}{c}{\textit{Metallicity tests}}  &   \multicolumn{3}{c}{\textit{SFH tests}} &   \multicolumn{2}{c}{\textit{Dust tests}}\\
\hline
$\log M_{\star}$ &   $\log\phi(M_{\star})$ &   $\Delta\log\phi$    &   $\Delta\log\phi$    &   $\Delta\log\phi$   &   $\Delta\log\phi$  &   $\Delta\log\phi$   &   $\Delta\log\phi$    &   $\Delta\log\phi$            &   $\Delta\log\phi$\\
                                  &                         &                       &   (0.01\,Z$_{\sun}$)  &   (0.5\,Z$_{\sun}$)  &   (delayed)         &   (rising)           &   (declining)         &   ($\hat{\tau}_V\in[0,3]$)    &   (SMC law)\\
M$_{\sun}$  &   Mpc$^{-3}$M$_{\sun}^{-1}$   &   Mpc$^{-3}$M$_{\sun}^{-1}$   &   Mpc$^{-3}$M$_{\sun}^{-1}$   &   Mpc$^{-3}$M$_{\sun}^{-1}$   &   Mpc$^{-3}$M$_{\sun}^{-1}$   &   Mpc$^{-3}$M$_{\sun}^{-1}$   &   Mpc$^{-3}$M$_{\sun}^{-1}$   &   Mpc$^{-3}$M$_{\sun}^{-1}$   &   Mpc$^{-3}$M$_{\sun}^{-1}$\\
\hline
$8.43$  &   $-2.42$ &   $0.11$  &   $0.11$  &   $0.11$  &   $0.11$  &   $0.11$  &   $0.11$  &   $0.15$  &   $0.11$\\
$8.24$  &   $-2.22$ &   $0.10$  &   $0.10$  &   $0.10$  &   $0.11$  &   $0.10$  &   $0.10$  &   $0.11$  &   $0.11$\\
$8.05$  &   $-1.98$ &   $0.09$  &   $0.09$  &   $0.09$  &   $0.10$  &   $0.09$  &   $0.12$  &   $0.11$  &   $0.10$\\
$7.86$  &   $-1.81$ &   $0.21$  &   $0.21$  &   $0.21$  &   $0.20$  &   $0.21$  &   $0.20$  &   $0.21$  &   $0.21$\\
$7.67$  &   $-1.71$ &   $0.16$  &   $0.17$  &   $0.16$  &   $0.16$  &   $0.16$  &   $0.16$  &   $0.16$  &   $0.16$\\
$7.48$  &   $-1.54$ &   $0.37$  &   $0.37$  &   $0.38$  &   $0.37$  &   $0.36$  &   $0.37$  &   $0.37$  &   $0.37$\\
$7.29$  &   $-1.36$ &   $0.39$  &   $0.39$  &   $0.38$  &   $0.38$  &   $0.40$  &   $0.38$  &   $0.39$  &   $0.39$\\
$7.10$  &   $-1.39$ &   $0.74$  &   $0.74$  &   $0.75$  &   $0.74$  &   $0.72$  &   $0.73$  &   $0.72$  &   $0.72$\\
$6.91$  &   $-1.38$ &   $0.56$  &   $0.57$  &   $0.56$  &   $0.62$  &   $0.56$  &   $0.57$  &   $0.54$  &   $0.56$\\
$6.72$  &   $-1.47$ &   $0.75$  &   $0.74$  &   $0.76$  &   $0.74$  &   $0.77$  &   $0.85$  &   $0.74$  &   $0.75$\\
$6.53$  &   $-1.56$ &   $1.41$  &   $1.60$  &   $1.41$  &   $1.42$  &   $1.41$  &   $1.41$  &   $1.42$  &   $1.42$\\
\hline
\end{tabular}
\end{table*}

\begin{figure}
    \centering
    \includegraphics[width=0.5\textwidth, keepaspectratio=True]{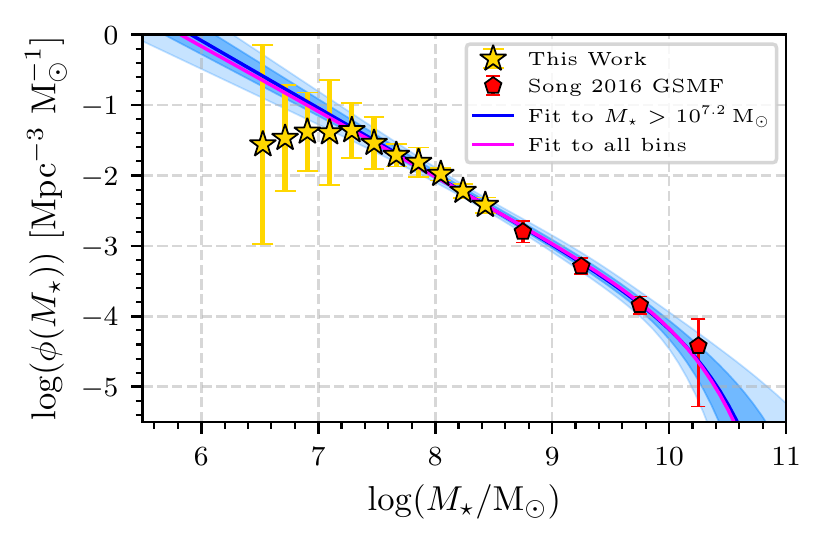}
    \caption{The same graph as in the \textit{left-hand} panel of Fig.~\ref{fig:GSMF-data-fit} with in addition a 'classical' Schechter fit to all mass bins, including those $\log(M_{\star}/\mathrm{M}_{\sun})<7.2$, as the magenta curve. It shows that including the lowest-mass bins does not significantly affect the best-fit Schechter function in the reference model.}
    \label{fig:GSMF_classical_schechter_full}
\end{figure}

\begin{figure*}
    \centering
    \includegraphics[width=\textwidth, keepaspectratio=True]{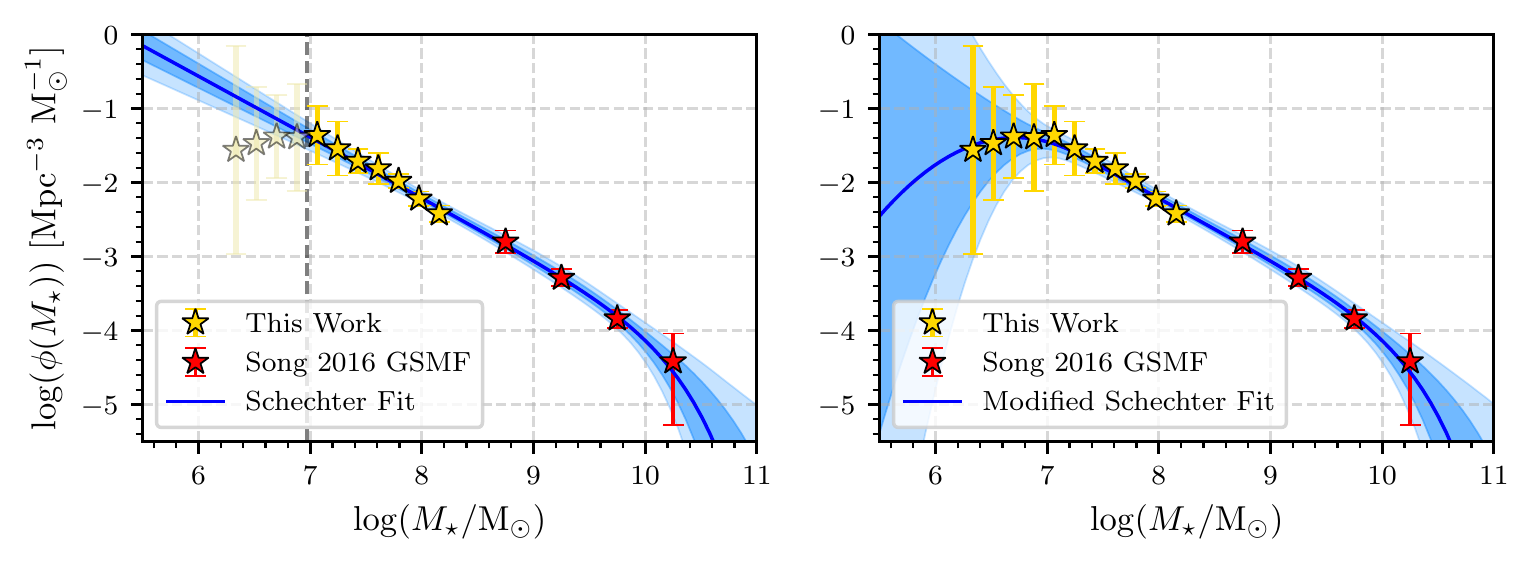}
    \caption{The same graphs as in Fig.~\ref{fig:GSMF-data-fit} but for the exponentially rising SFH model ($\psi(t)\propto\exp{(t/\tau)}$). This configuration allows for an upward turnover of the GSMF at the $1\sigma$-level (blue shaded area in the right panel).}
    \label{fig:GSMF-rising_SFH}
\end{figure*}

\begin{figure*}
    \centering
    \includegraphics{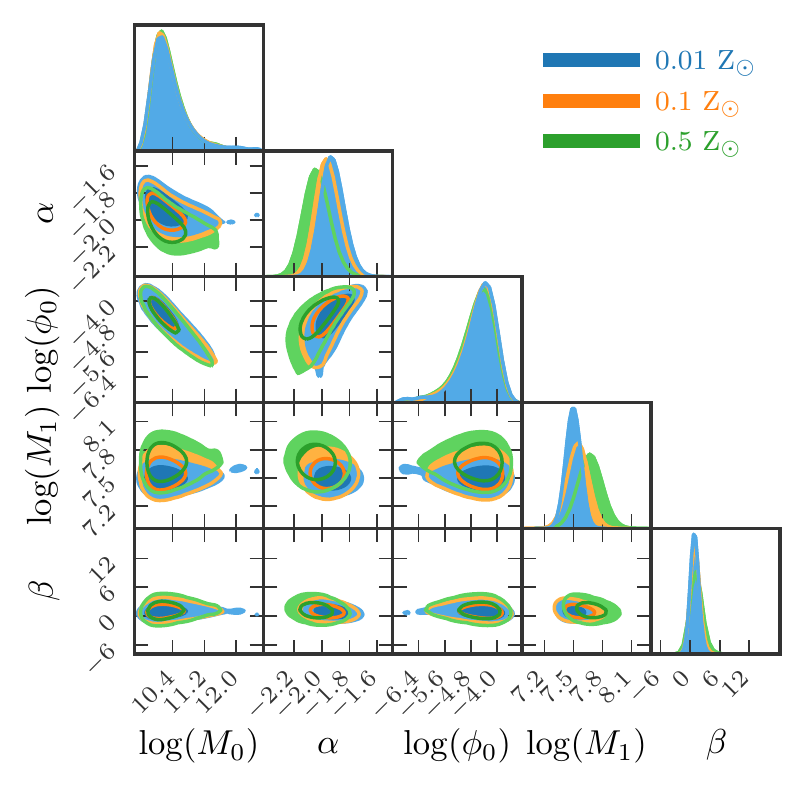}
    \includegraphics{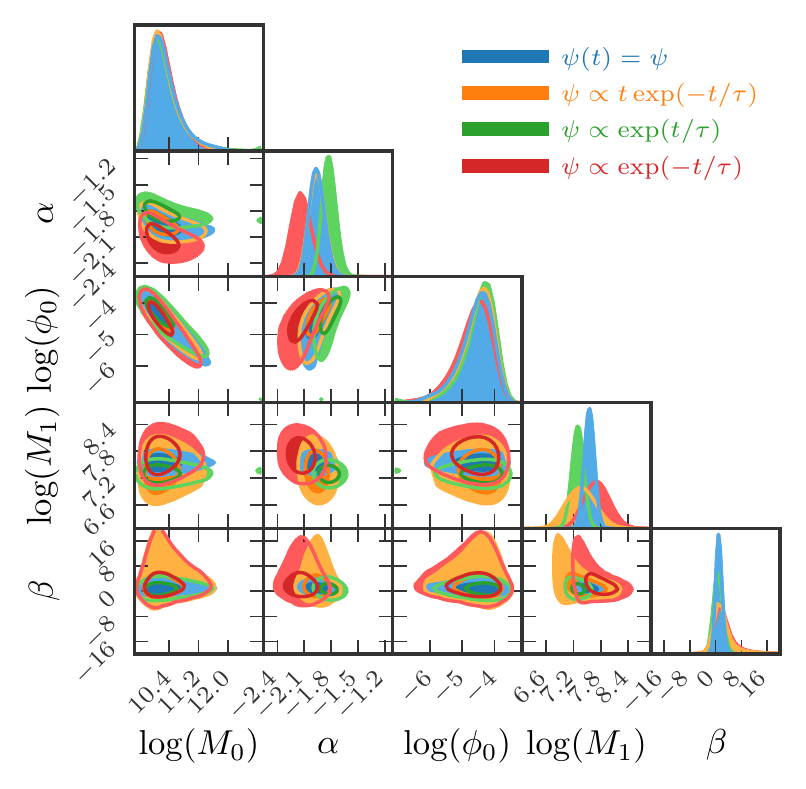}
    \includegraphics{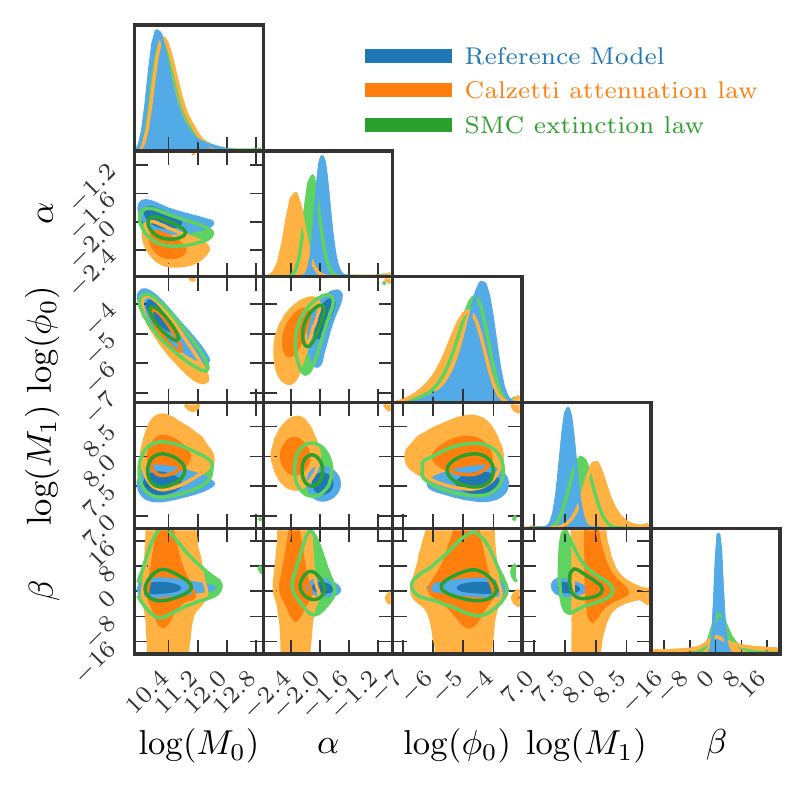}
    \caption{\textit{Upper right panel}: Posterior distributions of the four modified Schechter parameters for assuming different metallicities in the SED-fit. The resulting GSMF is not strongly impacted by metallicity. \textit{Upper left panel}: Posterior distributions of the four modified Schechter parameters for assuming different SFHs in the SED-fit. The most notable impact of SFH is on the low-mass end slope $\alpha$: A rising SFH results in a shallower and a declining SFH in a steeper low-mass end slope. The constant and the delayed SFH on the other hand almost yield the same GSMF. \textit{Lower panel}: Posterior distributions of the four modified Schechter parameters for assuming different dust attenuation laws and values in the SED-fit: Calzetti law with $\hat{\tau}_V\in[0,0.2]$ (blue), Calzetti law with $\hat{\tau}_V\in[0,3]$ (orange) and SMC law with $\hat{\tau}_V\in[0,3]$ (green). Dust attenuation mostly impacts the low-mass end of the GSMF, resulting in steeper low-mass end slopes $\alpha$ and in much larger uncertainties on the curvature $\beta$ when allowing $\hat{\tau}_V\in[0,3]$.}
    \label{fig:corner_plots}
\end{figure*}

In this section we provide additional material to the GSMFs computed in section~\ref{sec:GSMF}. Table~\ref{tab:GSMF_points} shows the stellar mass values of our GSMF points for each \texttt{BEAGLE} configuration from Table~\ref{tab:SED-fit_parameters}, computed from the rest-frame UV luminosity bins in \citet{atek18}. The four wide-area blank field high-mass GSMF points that we use in our GSMF-fits can be found in Table~2 in \citet{song16}. The $\phi(M_{\star})$ of our GSMFs are then shown in Table~\ref{tab:GSMF_errors} along with their total errors for each \texttt{BEAGLE} configuration. These errors combine gravitational lensing and rest-frame UV photometry uncertainties from \citet{atek18}, cosmic variance uncertainty by \citet{robertson14} and uncertainties in the determination of stellar mass from our SED-fitting analysis including the uncertainty in the empirical correction factor derived in section~\ref{sec:IRAC-photometry_correction}. Note that the errors do not strongly differ between the different SED-fitting models which indicates them to be dominated by the lensing uncertainties.

We show in Fig.~\ref{fig:GSMF_classical_schechter_full} that including the lowest-mass bins ($\log(M_{\star}/\mathrm{M}_{\sun})<7.2$) in the the 'classical' Schechter function fit does not significantly affect the best-fitting parameters in the reference model. Figure~\ref{fig:corner_plots} presents the posterior distributions for the four modified Schechter parameters of the GSMF fits for each SED-fitting model. As mentioned in section~\ref{sec:GSMF}, the GSMF does not strongly depend on metallicity (upper left panel). The posterior distributions of the GSMF fit parameters for the SFH and dust attenuation tests (upper right and lower panels in Fig.~\ref{fig:corner_plots} respectively) further illustrate the strong impacts of SFH and dust attenuation on the GSMF, in particular its low-mass end slope $\alpha$. They also emphasize the correlation between $\alpha$ and the low-mass end turnover curvature $\beta$. The upper right and lower panels of Fig.~\ref{fig:corner_plots} in particular illustrate that the two models with the steepest low-mass end slopes $\alpha$ allow for a much wider range in $\beta$-space, resulting in the very large uncertainties in $\beta$ discussed in section~\ref{sec:GSMF}.

Apart from those two cases where $\beta$ is very poorly constrained, only in the exponentially rising SFH case the MCMC fit of the GSMF allows for an upward turnover in the $1\sigma$-range (blue shaded area in Fig.~\ref{fig:GSMF-rising_SFH}) with a well constrained curvature. This is also the model that yields the shallowest low-mass end slope in our study with $\alpha\simeq-1.82_{-0.07}^{+0.08}$. Our modified Schechter parametrization of the GSMF, cf. Eq.~\eqref{eq:modified_schechter}, and the MCMC-fitting procedure therefore seem to disfavor $\beta<0$ solutions for our measured GSMF points for steeper low-mass end slopes. This enables us to place a lower boundary on the range of low-mass end slopes that allow for an upward turnover at the $1\sigma$ level at $\alpha\gtrsim-1.82$ (cf. Table~\ref{tab:schechter-parameters}). Since the shallower $\alpha$ in the exponentially rising SFH case is a direct consequence of overall lower stellar masses, as discussed in section~\ref{sec:SFH-impact}, this implies that $\beta\leq0$ (i.e. the GSMF does not turn over or has an upward turnover) requires the majority of galaxies at $z\sim6-7$ to be very very young, very low-mass galaxies with high exponentially increasing SFRs. Any conclusion of this sort would however require a much more robust derivation of the extremely low-mass end of the $z\sim6-7$ GSMF.


\bsp	
\label{lastpage}
\end{document}